\documentclass[%
reprint, hidelinks,
superscriptaddress,
hypertext,
showkeys,
showpacs,
 nofootinbib,
 nobibnotes,
 amsmath,amssymb,
 aps,
prc,
floatfix,
]{revtex4-1}

\usepackage{graphicx}% Include figure files
\usepackage{dcolumn}% Align table columns on decimal point
\usepackage{bm}% bold math
\usepackage{isotope}
\usepackage{upgreek}
\usepackage{longtable}
\usepackage[caption=false]{subfig}
\usepackage{booktabs}
\usepackage{threeparttable}
\usepackage{epigraph}
\usepackage{comment}
\usepackage[]{hyperref}% add hypertext capabilities
\hypersetup{
  colorlinks   = true, %Colours links instead of ugly boxes
  urlcolor     = blue, %Colour for external hyperlinks
  linkcolor    = blue, %Colour of internal links
  citecolor   = blue %Colour of citations
}
\usepackage[mathlines]{lineno}% Enable numbering of text and display math
\newcommand\Tstrut{\rule{0pt}{3.0ex}}       % top strut
\begin{document}

\preprint{APS/123-QED}

\title{Cross--section measurements of radiative proton--capture reactions\\
in \isotope[112]{Cd} at energies of astrophysical interest}

\author{A.~Psaltis}
\altaffiliation[Present address: ]{
 Department of Physics and Astronomy,
 McMaster University,
 Hamilton, ON L8S 4M1,
 Canada}%\email{psaltisa@mcmaster.ca} 
\affiliation{Department of Physics,
National Kapodistrian University of Athens,
Zografou Campus,
GR--15784,
Athens, Greece
}%

\author{A.~Khaliel}%
\affiliation{Department of Physics,
National Kapodistrian University of Athens,
Zografou Campus,
GR--15784,
Athens, Greece
}%

\author{E.--M.~Assimakopoulou}%
 \altaffiliation[Present address: ]{
 Department of Physics and Astronomy,
 Uppsala, Sweden}
\affiliation{Department of Physics,
National Kapodistrian University of Athens,
Zografou Campus,
GR--15784,
Athens, Greece
}%
 
\author{A.~Kanellakopoulos}%
 \altaffiliation[Present address: ]{
 Instituut voor Kern-- en Stralingsfysica,
 KU Leuven, Celestijnenlaan 200D,
 B--3001 Leuven,
 Belgium}
\affiliation{Department of Physics,
National Kapodistrian University of Athens,
Zografou Campus,
GR--15784,
Athens, Greece
}%
 
\author{V.~Lagaki}%
 \altaffiliation[Present address: ]{
CERN, Geneva, Switzerland}
\affiliation{Department of Physics,
National Kapodistrian University of Athens,
Zografou Campus,
GR--15784,
Athens, Greece
}%

\author{M.~Lykiardopoulou}%
 \altaffiliation[Present address: ]{
 Department of Physics and Astronomy,
 University of British Columbia,
 6224 Agricultural Road,
 Vancouver, BC V6T1Z1,
 Canada}
\affiliation{Department of Physics,
National Kapodistrian University of Athens,
Zografou Campus,
GR--15784,
Athens, Greece
}%
 
\author{E.~Malami}%
 \altaffiliation[Present address: ]{
 Nikhef, Science Park 105,
 1098 XG, Amsterdam,
 Netherlands}
\affiliation{Department of Physics,
National Kapodistrian University of Athens,
Zografou Campus,
GR--15784,
Athens, Greece
}%

\author{P.~Tsavalas}%%
\affiliation{Department of Physics,
National Kapodistrian University of Athens,
Zografou Campus,
GR--15784,
Athens, Greece
}%
\affiliation{
INRASTES, NCSR ``Demokritos'', GR--15310, Aghia Paraskevi, Greece
}%

\author{A.~Zyriliou}%
\affiliation{Department of Physics,
National Kapodistrian University of Athens,
Zografou Campus,
GR--15784,
Athens, Greece
}%

\author{T.J.~Mertzimekis}%
\email{Corresponding author: tmertzi@phys.uoa.gr}
\affiliation{Department of Physics,
National Kapodistrian University of Athens,
Zografou Campus,
GR--15784,
Athens, Greece
}%

\date{\today}% It is always \today, today,
             %  but any date may be explicitly specified

\begin{abstract}
Reactions involving the group of nuclei commonly known as \textit{p} nuclei
are part of the nucleosynthetic mechanisms at astrophysical sites.
The \isotope[113]{In} nucleus is such a case with several open questions
regarding its origin at extreme stellar environments. In this work, the
experimental study of the cross sections of the radiative proton--capture
reaction \isotope[112]{Cd}($p,\gamma$)\isotope[113]{In} is attempted for the
first time at energies lying inside the Gamow window with an isotopically
enriched \isotope[112]{Cd} target. Two different techniques, the in--beam
$\gamma$--ray spectroscopy and the activation method, have been applied.
The latter method is required to account for the presence of a low--lying
\isotope[113]{In} isomer at 392 keV having a halflife of $\approx 100$~min.
From the cross sections, the astrophysical \textit{S} factors and the isomeric
ratios have been additionally deduced. The experimental results are compared
to detailed Hauser--Feshbach theoretical calculations using \texttt{TALYS}, and
discussed in terms of their significance to the optical model potential involved.

\begin{comment}
The production of the \textit{p}--nuclei remains an open question
for nuclear astrophysics. It is based on a huge reaction network,
and their final abundances are theoretically calculated using the
Hauser--Feshbach statistical model, which relies on scarce experimental
data. In addition, the origin of \isotope[113]{In} presents a
long--standing puzzle, since it is underproduced in most
nucleosynthesis models. Towards that direction, a campaign of
measurements at the Tandem Accelerator Laboratory of NCSR
``Demokritos'', focusing on total reaction cross sections of the
\isotope[112]{Cd}($p,\gamma$)\isotope[113]{In} reaction was
undertaken. The reaction was studied using a set of high--purity
germanium detectors by in--beam $\gamma$--ray spectroscopy to study
prompt $\gamma$--rays and the activation method for an existing
low--lying isomeric state of \isotope[113]{In} (E=$391.7$~keV,
$t_{1/2}=99.5$~min). Total cross sections for four proton beam
energies inside the Gamow window for temperatures $T=2-3$~GK were
measured for the first time. Comparison between experimental results
and the Hauser--Feshbach theoretical calculations with the
\texttt{TALYS} code shows a fairly good agreement. The experimental
results provide new input for the theoretical modeling of the
\textit{p}--process in this mass region near \isotope[113]{In}.
\end{comment}

\end{abstract}


\maketitle

%\tableofcontents

\section{Introduction}
\label{sec:intro}

The origin of some 35 neutron--deficient stable isotopes with mass 
$A\geq74$, between \isotope[74]{Se} and \isotope[196]{Hg}, in the
neutron--deficient side of the valley of stability, commonly known as
``\textit{p} nuclei'', has been one of the major open questions in nuclear
astrophysics~\cite{burbidge1957synthesis,cameron1957nuclear}. The solar
abundances of \textit{p} nuclei are one to two orders of magnitude lower
compared to the respective \textit{r} and \textit{s} nuclides in the
same mass region~\cite{lodders20094}, which is attributed to ``shielding''
by their reaction flow~\cite{arnould2003p, rauscher2013constraining}.

Various astrophysical environments and associated processes have been
proposed to explain the origin of the \textit{p} nuclei and their solar
abundances. The main mechanism is referred to as the \textit{p} process,
but it is used interchangeably with the term $\gamma$ process, which also
plays a dominant role to this nucleosynthesis scenario~\cite{woosley1978p}.
The \textit{p} process is assumed to occur in different zones inside
a core--collapse supernovae, and thus the peak temperature for the
\textit{p} process lies between $T_{peak}\sim 2-3$~GK~\cite{arnould2003p}.
It has also been shown that the \textit{p} process can also occur in a
single--degenerate type Ia supernovae scenario~\cite{travaglio2011type}.

Several other explosive nucleosynthesis scenarios, such as the \textit{rp}
process~\cite{schatz1999rapid}, the \textit{pn} process~\cite{goriely2002he}
and the $\nu$\textit{p}
process~\cite{frohlich2006neutrino, pruet2006nucleosynthesis, wanajo2006rp}
have been proposed to contribute to the production of \textit{p} nuclei.
It is remarkable that despite the variety of astrophysical models,
these processes can reproduce the solar abundances of the \textit{p} nuclei
within a factor of 3 (e.g. see the sensitivity study by Rapp
\textit{et al.}~\cite{rapp2006sensitivity}). Nevertheless, several species,
such as \isotope[92,94]{Mo}, \isotope[96,98]{Ru}, \isotope[113]{In}
and \isotope[115]{Sn}, are significantly underproduced in most models.
In the context of the present work, the origin of \isotope[113]{In} is
discussed in some detail later in the text.

The vast \textit{p} process reaction network involves roughly $20\,000$
reactions among $2\,000$ nuclei~\cite{arnould2003p}
and thus, within that framework, most of the reaction rates need to be
estimated using the Hauser--Feshbach statistical model~\cite{hauser1952inelastic}.

The experimental input is invaluable in terms of constraining the model
parameters. Measurements of cross sections in radiative proton--capture
reactions can play a two--fold pivotal role towards the understanding
of the \textit{p} process. First, they can be used to adjust the parameters
of the statistical model improving theoretical predictions for
currently unmeasured reactions, and second, they can make calculations of
important photodisintegration decay constants possible~\cite{jose2011nuclear}.

\subsection*{Open questions on the origin of \isotope[113]{In}}
\label{sec:origin}

The production of \isotope[113]{In} at astrophysical sites has been a
long--standing puzzle for nuclear astrophysics~\cite{ward1981origin}.
\isotope[113][]{In} is the lightest in a group of four \textit{p} nuclei
that are not even--even\footnote{The other three are
\isotope[115]{Sn},
\isotope[138]{La} and
\isotope[180m]{Ta}$^{m}$.
\isotope[138]{La} is considered to be produced by the
$\nu$ process ($\nu$ flows from core--collapse
supernovae)~\cite{woosley1990nu}.}~\cite{rauscher2013constraining},
and has a relatively high elemental contribution of 4.3\%~\cite{lodders20094}.

\begin{figure*}[!ht]
\centering
\includegraphics[width= 0.9\textwidth]{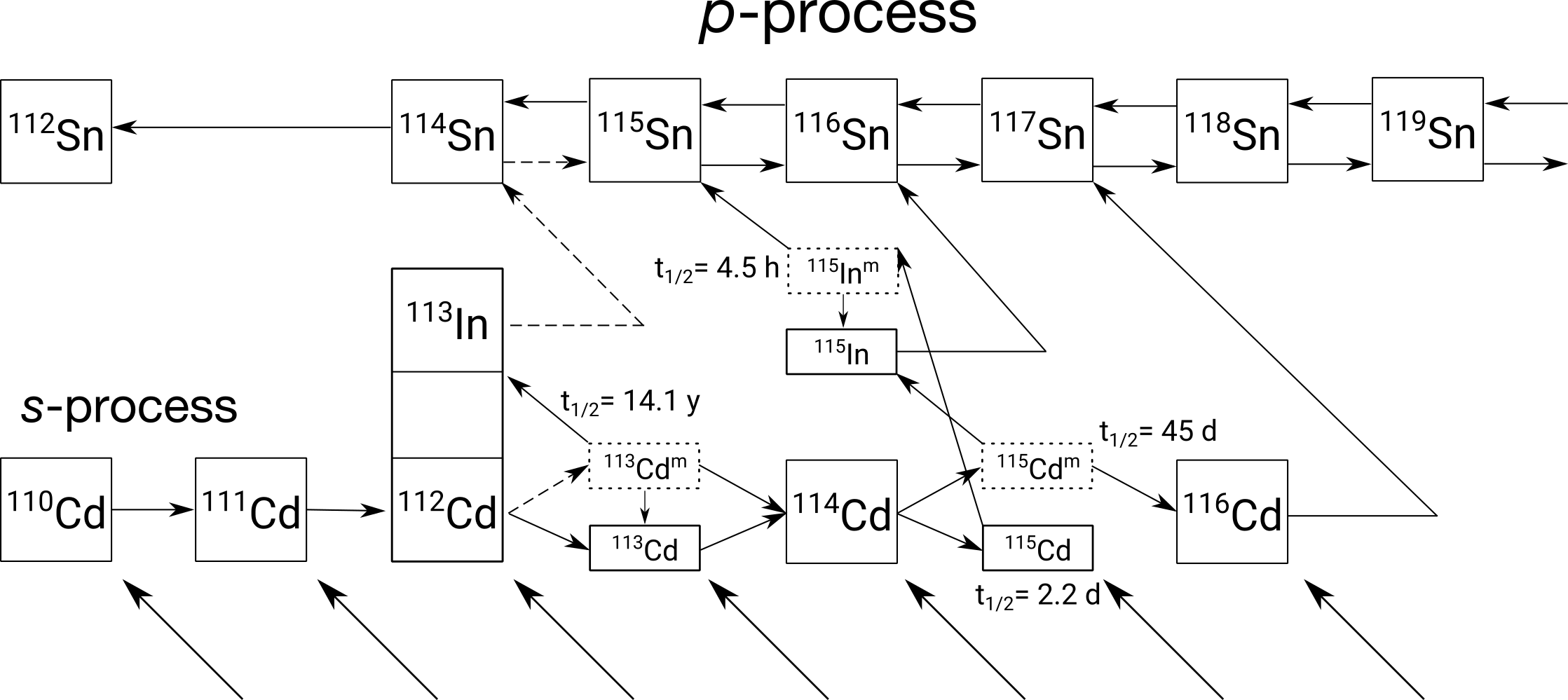}
\caption{A sketch of the reaction flows in the vicinity of \isotope[113]{In}
adapted from Ref.~\cite{hayakawa2016measurement} taking into account \cite{2006_Rauscher}.
Contributions from the corresponding \textit{s}, \textit{r} and \textit{p}
processes are shown. The present work focuses on the proton capture channel
by \isotope[112]{Cd}, which is marked in the figure with the strong--line box.
See text for details.
}
\label{fig:cdinsn}
\end{figure*}

The complexity of nucleosynthesis in the Cd--In--Sn region arises mainly
due to the existence of several long--lived $\beta$--decaying
isomers~\cite{nemeth1994nucleosynthesis, theis1998puzzling} (see also
Fig.~\ref{fig:cdinsn}) and leads to significant underproduction of the
rare odd--A isotopes \isotope[113]{In} and \isotope[115]{Sn}~\cite{rapp2006sensitivity}.

Nemeth \textit{et al.}~\cite{nemeth1994nucleosynthesis} proposed a
\textit{s}--process contribution to the origin of \isotope[113]{In},
which was calculated to be very small (less than 1\%). Recent calculations,
using KADoNiS~\cite{dillmann2006kadonis} have resulted in a much smaller,
0.0013\% contribution.                                                            

Theis \textit{et al.} have showed that post--\textit{r} process
$\beta$--decay chains could account for less than 12\% of the solar
abundance of \isotope[113]{In}, and that thermally enhanced
$\beta$ decay of the progenitor \isotope[114]{Cd} is possible~\cite{theis1998puzzling}.
Finally, Dillmann \textit{et al.}~\cite{dillmann2007p, dillmann2008there}
proposed the $\beta$--delayed \textit{r} process decay chains as the most
promising scenario.

The \textit{rp} and $\nu$\textit{p} processes are excluded as possible
production mechanisms, since they generally produce nuclei up to
$A=110$~\cite{dillmann2008there}. In this context, a $\nu$\textit{p} process
sensitivity study by Wanajo~\textit{et al.}~\cite{wanajo2011uncertainties}
has demonstrated that by changing either astrophysical or nuclear physics
input parameters, the $\nu$\textit{p} process could account for the origin
of \isotope[113]{In} and other $A>110$ \textit{p} nuclei.

Concerning possible astrophysical sites, Fujimoto \textit{et al.} showed
in Ref.~\cite{fujimoto2007heavy} that \isotope[113]{In} and several other
underproduced \textit{p} nuclei can be abundantly synthesized in ejecta
originated by a collapsar~\cite{woosley1993gamma}. Specifically, the heavy
\textit{p} nuclei, including \isotope[113]{In}, are produced in the jets
through fission~\cite{fujimoto2007heavy}. 

Interestingly enough, it has been demonstrated by Babishov and
Kopytin~\cite{babishov2006model,kopytin2013role} that \isotope[113]{In}
could be produced during a supernova explosion of a $25M_{\odot}$ star.
However, their final \textit{p} abundances are accompanied by underestimated
molybdenum and ruthenium abundances, still leaving some open questions.

As a consequence of all the above, it is nowadays widely accepted
that \isotope[113]{In} is not a ``pure'' \textit{p} nucleus, but
has non--negligible contributions from the \textit{s} and \textit{r}
processes~\cite{pignatari2016production}.

Many studies have focused on \isotope[113]{In} in the vicinity of
$\gamma$--process nucleosynthesis energies, such as
the \isotope[113,115]{In}($p,\gamma$)\isotope[114,116]{Sn} reactions~\cite{harisso113In},
the $\alpha$ elastic scattering~\cite{kiss2013high}, and
the \isotope[113]{In}($\alpha,\gamma$)\isotope[117]{Sb} reactions~\cite{yalccin2009odd}.
Recently, Muhammed Shan \textit{et al.}~\cite{2018_MuhammedShan}
focused on proton--induced reactions in \isotope[113]{In} at energies
ranging 8--22~MeV adding information to earlier investigations of the
\isotope[112]{Cd}($p,n$)\isotope[113]{In} reaction~\cite{Blaser_1951,NSR1975AB09,Skakun_1975}. The spin isomer
in \isotope[113]{In} was also very recently studied in the pygmy
resonance region with photoexcitation~\cite{2019_Nedorezov}.

In the present work, we report on a first experimental attempt to study
the radiative proton capture relevant to the production of
\isotope[113]{In} by measuring the reaction cross sections at
astrophysically interesting energies, using an isotopically enriched 
\isotope[112]{Cd} target. Despite the particular reaction is not
necessarily a strong channel in the reaction flow~\cite{2006_Rauscher},
it can still be considered valuable to have its cross section measured,
as it can assist in constraining models to offer better predictions
for reactions that can not be measured directly in this mass regime.

\section{Experimental Details}
\label{sec:exp}

Measurements for the study of the radiative proton capture reaction on
\isotope[112]{Cd} were carried out at the 5.5 MV T11 Tandem Van de Graaff
accelerator of the NCSR ``Demokritos'' in Athens, Greece. Both the in--beam
and activation methods have been used in the measurements to account for
a low--lying isomeric state in the populated nucleus \isotope[113]{In}.

\subsection{The Proton Beams}
\label{sec:beam}

The reaction \isotope[112]{Cd}($p,\gamma$)\isotope[113]{In}
($Q=6081.2(2)$~keV)~\cite{qvalue} was studied at four proton lab
energies in total, i.e. 2.8, 3.0, 3.2 and 3.4~MeV. All energies lie
inside the Gamow Window for temperatures related to the production
of \textit{p} nuclei with $A\sim 92-144$ at $T_{peak}=2-3$~GK, which
corresponds to $E_p=1.8-4.5$~MeV. During the
experiments the target was irradiated with protons of beam currents
ranging $150-300$~enA.

\subsection{The Target}
\label{sec:tgt}

A multi--layer target was irradiated during the experiments, comprising
a front layer of 99.7\% enriched \isotope[112]{Cd} evaporated on a \isotope[nat]{Bi}
layer, backed by an \isotope[nat]{In} layer and a thick \isotope[nat]{Cu} layer.
Considering the generally low proton--capture cross section at these energies and
the low natural abundance of \isotope[112]{Cd}, the use of an enriched
target was imperative. The thick \isotope[nat]{Cu} backing provided
efficient charge collection during the experiment.
\begin{figure}[ht]
\centering
\includegraphics[width=.48\textwidth]{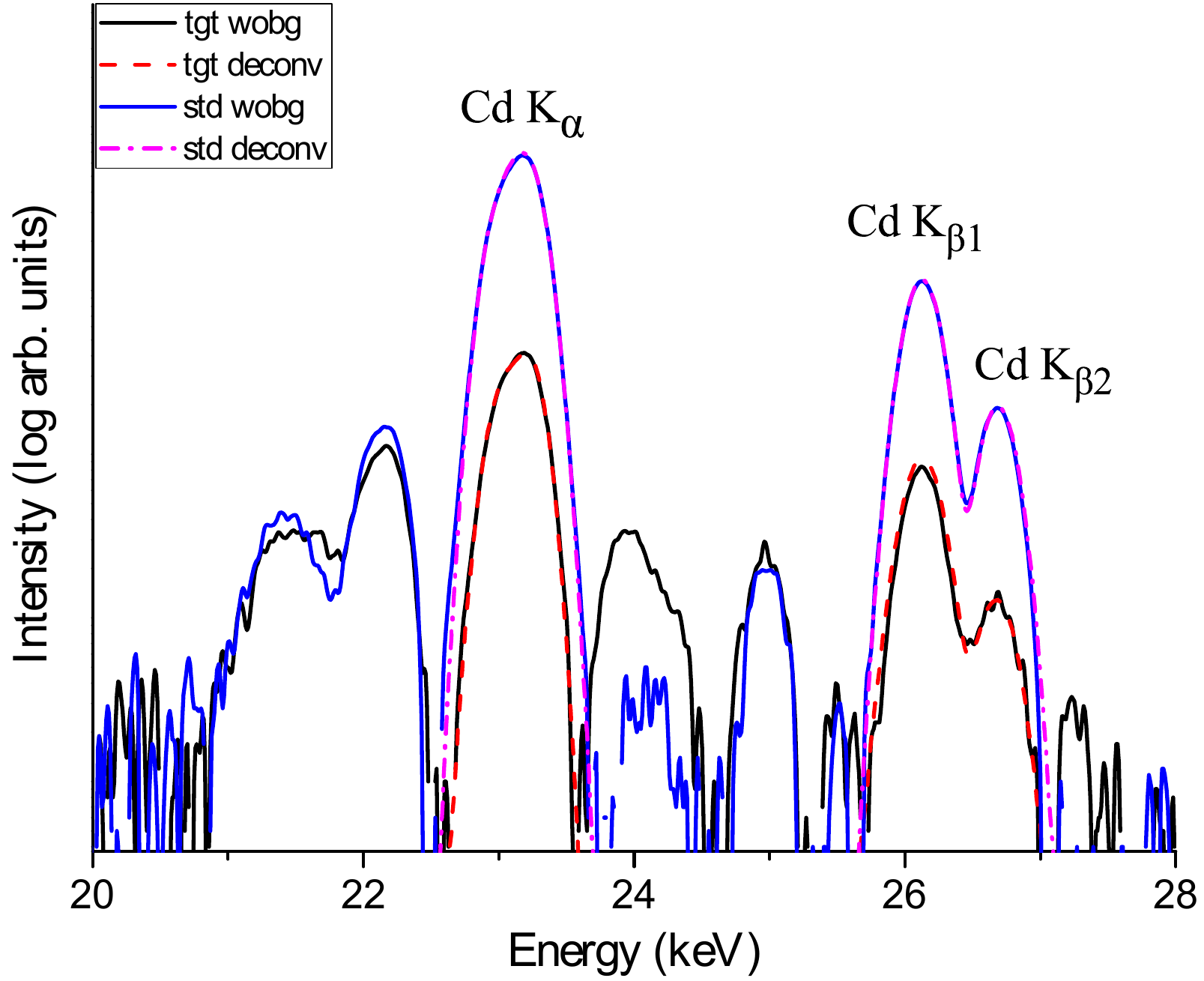}
\caption{The X--ray fluorescence spectrum of the target (\textit{tgt})
after background removal (\textit{wobg}) and photopeak deconvolution
(\textit{deconv}), as compared to a standard Cd sample (\textit{std}).
}
\label{fig:xrf}
\end{figure}

The \isotope[112]{Cd} layer thickness was measured equal to
$\delta_{\textrm{RBS}}= 0.96$~mg/cm$^2$ with the Rutherford
Backscattering Technique (RBS) before and after the experiment
and found to have no degradation due to irradiation~\cite{2019_Gyurky}.
To further confirm the layer thickness, an independent measurement
was carried out after the experiment using X--ray Fluorescence
Spectroscopy (XRF) resulting in a value of
$\delta_{\textrm{XRF}}=1.02$~mg/cm$^2$.
The two results were combined to produce the average value of
$\delta_{avg}=0.99(5)$~mg/cm$^2$, where the error cited is the standard
deviation calculated from the two measurements.

The target was turned inside the chamber by $30^\circ$ with respect
to the beam to avoid having its aluminum frame masking any of the
surrounding HPGe detectors, in particular the one sitting at 90$^\circ$
(see also Ref.~\cite{khaliel2017first}), thus
resulting in an effective thickness of the target,
$\delta= \frac{\delta_{avg}}{\cos 30^\circ} = 1.14(6)$~mg$\cdot$cm$^{-2}$.

Proton--beam energy losses in the target were calculated using
{SRIM}2013~\cite{ZIEGLER20101818} and found to be $\Delta E=59-52$~keV
for the corresponding proton beam energies $E_p=2.8-3.4$~MeV in the
laboratory frame. Assuming reactions taking place in the middle of the
\isotope[112]{Cd} layer, the effective energy in the center--of--mass
system is given by (see also Table~\ref{tab:gs_cross_sections}):
\begin{equation}
E_{eff}= E_p -\frac{\Delta E}{2}
\end{equation}

A voltage of $-300$~V was applied to the target chamber to suppress
the emission of secondary electrons from altering the charge collection
readings, which are essential for the calculation of the reaction yields
and subsequently the cross section. The target was mounted on an aluminum
heatsink cooled externally by an air--pumping system.

\subsection{Detection Apparatus \& Experimental Methods}

\begin{figure}[ht]
\centering
\includegraphics[width=0.47 \textwidth]{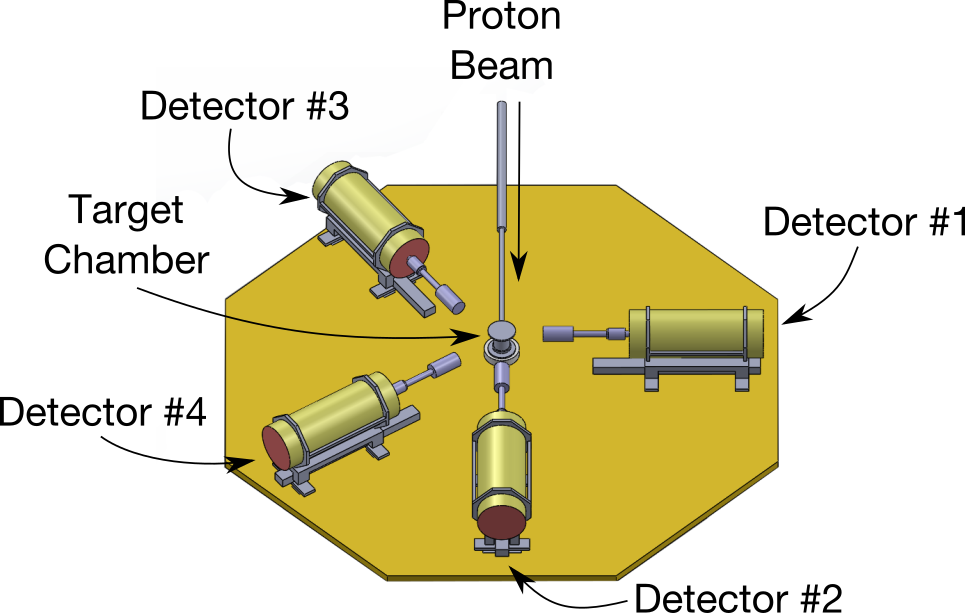}
\caption{A CAD model of the experimental setup used in the present
work. The target chamber was surrounded by an array of four HPGe detectors
placed on a turntable to measure $\gamma$--singles from eight different angles.
}
\label{fig:table}
\end{figure}

An array of four high--purity Germanium (HPGe) detectors of 100\% relative
efficiency was mounted on an octagonal turntable with maximum radius 2.4~m
(Fig.~\ref{fig:table}).
The table's turning ability enables measurements of a full angular
distribution. This particular setup is known of its versatility on measuring
cross sections and angular distributions of radiative capture reactions
relevant to the \textit{p} process. Similar studies can be found in Refs.~\cite{galanopoulos200388,sauerwein2012investigation,khaliel2017first}.

\begin{figure*}[ht]
\centering
\includegraphics[width=.98\textwidth]{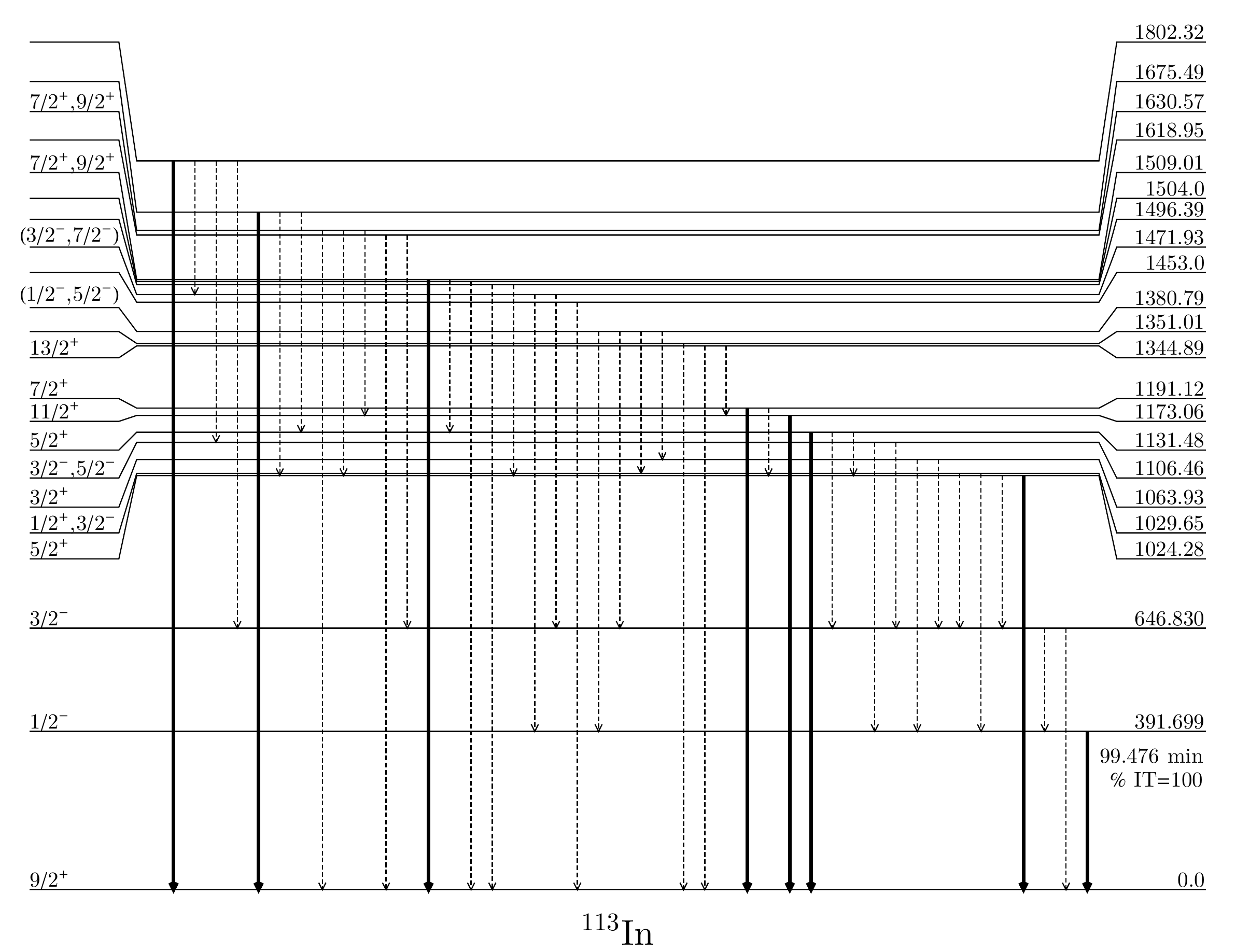}
\caption{A partial level scheme of the low--lying energy levels of
\isotope[113]{In}. Solid arrows represent decays feeding the ground state
of \isotope[113]{In} and were observed during our measurements. See transitions
marked with an asterisk in Fig.~\ref{fig:spectrum}.}
\label{fig:scheme}
\end{figure*}

\begin{figure}
\centering
\includegraphics[width=0.48\textwidth]{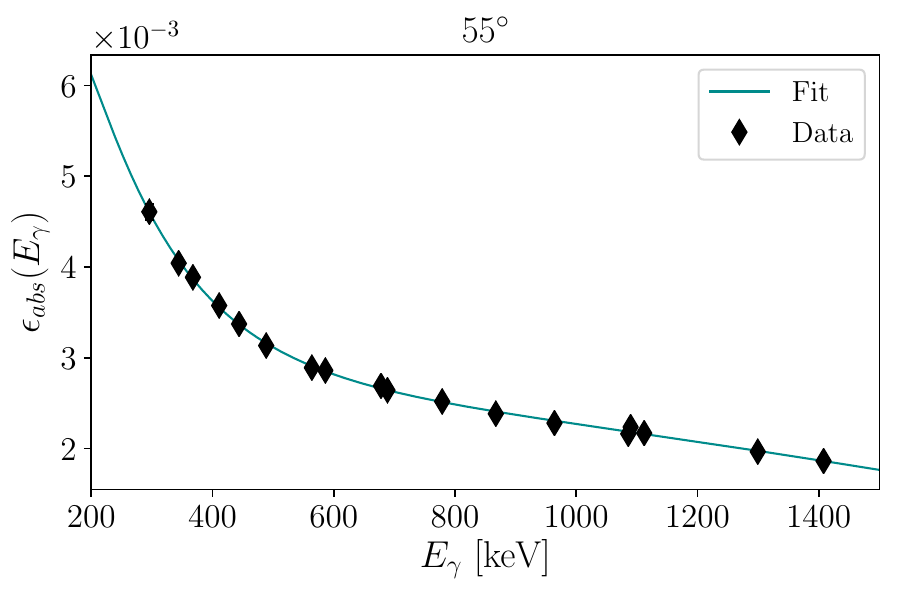}
\caption{A typical absolute efficiency curve for the detectors employed
in the measurement. The particular one corresponds to the detector
placed at 55$^\circ$. Errors are smaller than the symbol size.}
\label{fig:effic55}
\end{figure}

Detectors 1--4 were initially placed at 90, 0, 55 and 165 degrees,
respectively, with reference to the beam direction. Their distances from
the target were 15.5, 15.5, 14.8 and 18.0~cm, respectively. By turning
the table by 15 degrees counterclockwise, an additional set of angles was
used (105, 15, 40, 150 degrees respectively). Energy calibrations and absolute
efficiency measurements (Fig.~\ref{fig:effic55}) for all detectors were
performed with a standard \isotope[152]{Eu} point source placed in the
exact target position, before and after the experiments. Spectra were
recorded in singles mode using the nuclear electronics setup described
in Ref.~\cite{khaliel2017first}.

\begin{figure}
\centering
\includegraphics[width=0.46\textwidth]{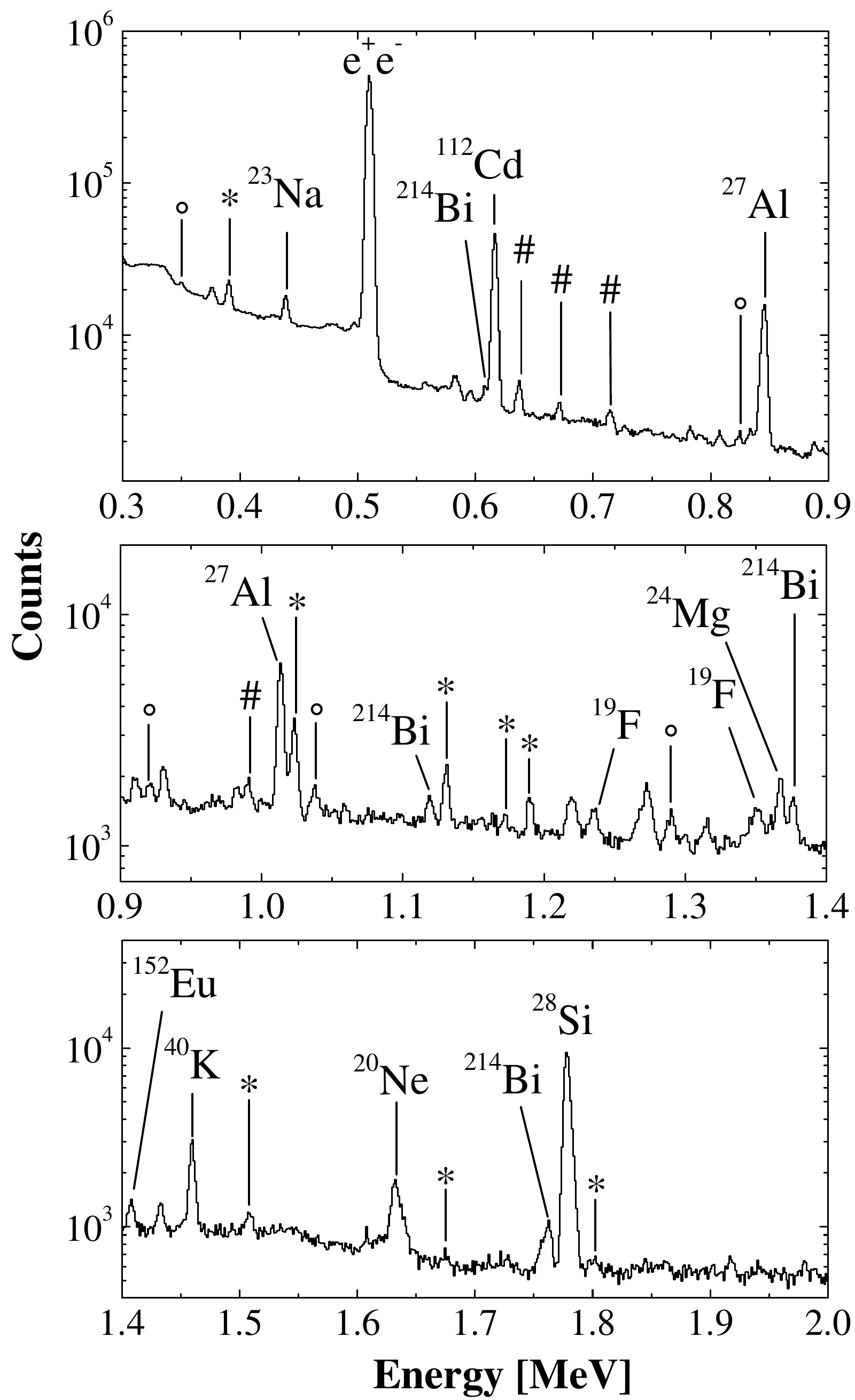}
\caption{A horizontal split--view (300--2000 keV) of a typical spectrum
recorded in singles in the detector placed at 55$^\circ$ and at a beam
energy of 3.4~MeV. Photopeaks feeding the ground state of \isotope[113]{In}
are marked with an $\ast$, whereas transitions feeding the isomeric $1/2^-$
state are marked with a \#. Other de-excitations of \isotope[113]{In} are
marked with circles. Major background lines which are usually observed in
the present setup, coming from natural radioactivity (e.g. \isotope[40]{K},
\isotope[214]{Bi}) or elements present in the beamline components (e.g.
\isotope[27]{Al}, \isotope[28]{Si}) are also labeled. Please note that
subfigure y--axes are not in scale.}
\label{fig:spectrum}
\end{figure}

Due to the structure properties of \isotope[113]{In} (see level scheme in
Fig.~\ref{fig:scheme}), two different methods were employed to study the
cross--section of the radiative proton capture reaction:
in--beam $\gamma$--ray spectroscopy, and activation. 

A low--lying isomeric state of \isotope[113]{In}
($E_{\gamma}$=391.7~keV, $t_{1/2}=99.476(23)$~min
(see Ref.~\cite{nndc} for the data and Fig.~\ref{fig:scheme}
for a partial level scheme) was populated in the reactions. Due to the
particular lifetime of the state, a measurement of the corresponding
cross section relies on the exploitation of the activation method.
In the recent past, similar studies have successfully employed the
activation technique~\cite{Yalcin2009, Kiss2011, Dillman2011,
Sauerwein2011, Halasz2012, Netterdon2013, Netterdon2014,
Guray2015, Kinoshita2016}.
For a more detailed description concerning the application of the
activation method on proton--induced reactions relevant to the
\textit{p} process, the reader is referred to Refs.~\cite{gyurky2003proton,2019_Gyurky}.

In the present case, the activation method was combined efficiently
with the in--beam measurements. The duration of irradiation was kept at
$\approx$ 6--8 h, to ensure that the isomeric state has been populated
sufficiently and (almost) reached saturation. Following irradiation,
overnight measurements for over five half--lives ($\approx 500$~min) were
performed, without beam delivery on the target. Activation measurements
followed in--beam measurements for each proton beam energy used in this
study.

\section{Data Analysis and Results}
\label{sec:analysis}

\subsection{In--beam measurements}

The cross section of the reaction \isotope[112]{Cd}($p,\gamma$)\isotope[113]{In}$_{gs}$
can be estimated from the relation~\cite{1988_Rolfs}:
\begin{equation}
\sigma_{gs} = \frac{A}{N_A} \frac{Y}{\delta}
\label{eq:xs}
\end{equation}
where $A$ is the atomic mass of the target in a.m.u.,
$N_A$ is the Avogadro number,
$\delta$ is the actual target thickness in $\mu g~cm^{-2}$ and
$Y$ is the absolute yield of the reaction in counts per $mC$.
The latter can be deduced from:
\begin{equation}
Y = \sum_i^n A_0^i
\label{eq:yield}
\end{equation}
where the $A_0^i$ coefficients are related to the angular distributions
of the emitted photons originating from the $i$--th $\gamma$ transition
feeding the ground state of the residual nucleus:
\begin{equation}
W^i(\theta)= A_0^i \left(1 + \sum_k \alpha_k^i P_k\left(\cos \theta \right) \right)
\ \text{for} \ k=2,4,\ldots
\label{eq:angdist}
\end{equation}
where the $a_k^i$ are coefficients which depend on the spin and parity
of the initial and final state of the transition, and $P_k$ are Legendre
polynomials. From the level scheme of the residual nucleus \isotope[113]{In}
(Fig.~\ref{fig:scheme}), seven transitions feeding the ground state were
observed with statistics above the background:

\begin{table}[ht]
    \begin{tabular}{rl}
    $ 5/2^+_1 \rightarrow 9/2^+_{gs}$ &~~E$_\gamma=1024$ keV\\
    $ 5/2^+_2 \rightarrow 9/2^+_{gs}$ &~~E$_\gamma=1132$ keV\\
    $ 11/2^+_2 \rightarrow 9/2^+_{gs}$ &~~E$_\gamma=1173$ keV\\
    $ 7/2^+_1 \rightarrow 9/2^+_{gs}$ &~~E$_\gamma=1191$ keV\\
    $ (7/2^+,9/2^+) \rightarrow 9/2^+_{gs}$ &~~E$_\gamma=1509$ keV\\
   unknown$ \rightarrow 9/2^+_{gs}$
&~~E$_\gamma=1676$ keV\\
   unknown$ \rightarrow 9/2^+_{gs}$
&~~E$_\gamma=1802$ keV\\
    \end{tabular}
\end{table}

\begin{figure*}[!ht]
\includegraphics[width=0.48\textwidth]{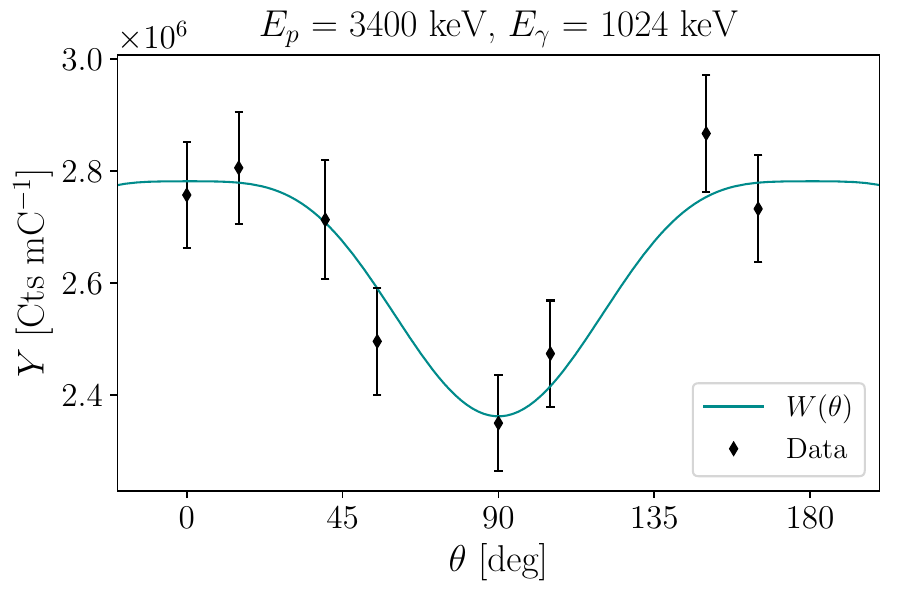}
\includegraphics[width=0.48\textwidth]{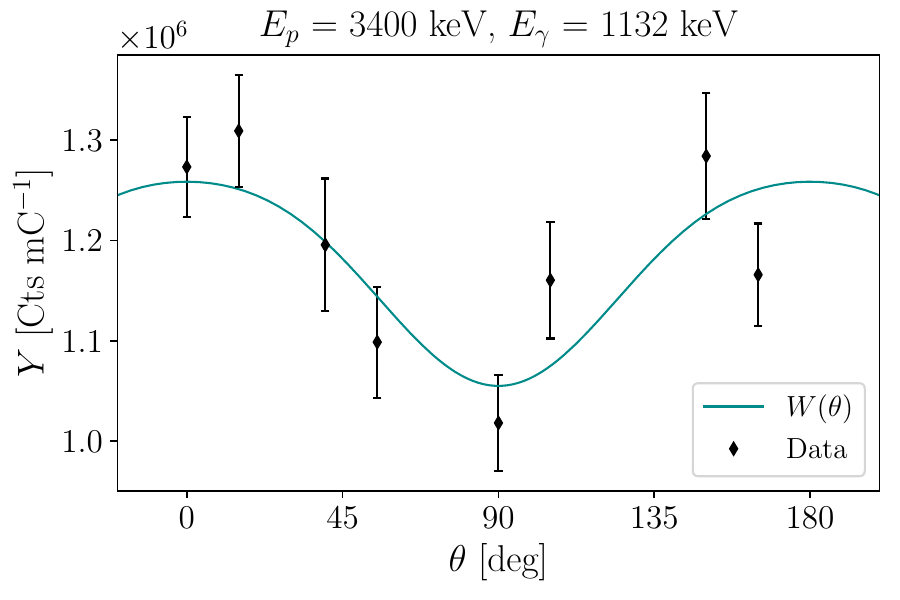}
\includegraphics[width=0.48\textwidth]{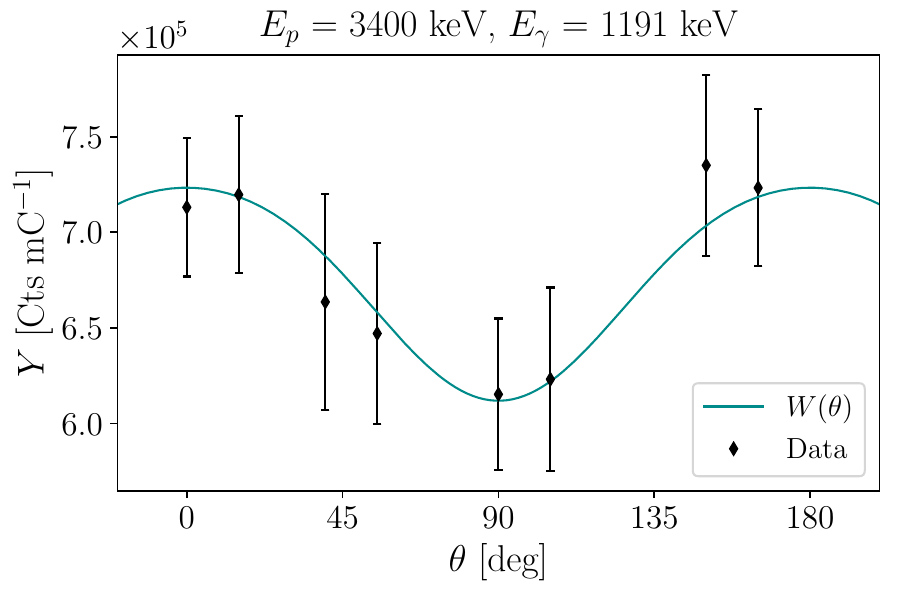}
\includegraphics[width=0.48\textwidth]{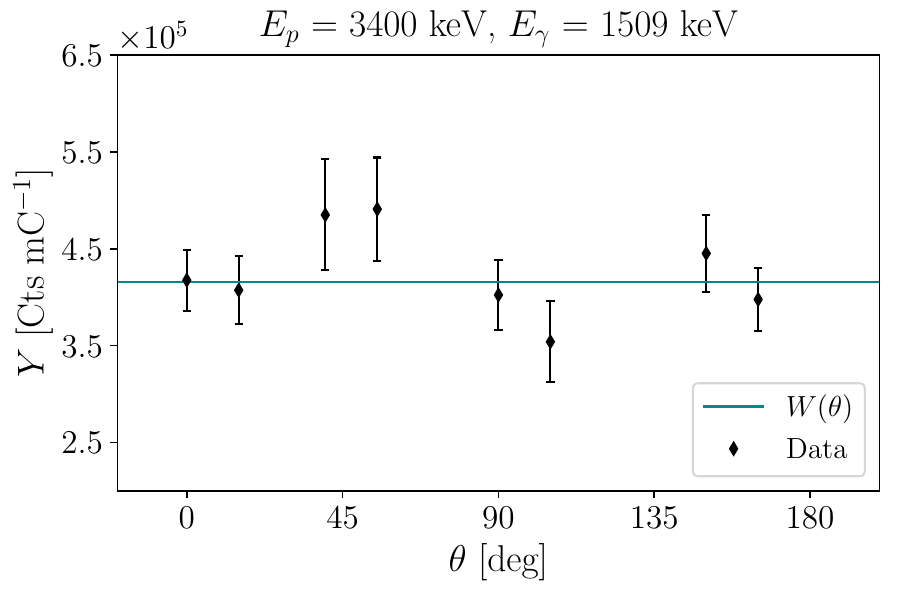}
\centering
\caption{Typical examples of angular distributions of the measured absolute yield
for the transitions
$5/2^+_1 \rightarrow 9/2^+_{gs}$ (top left),
$7/2^+_1 \rightarrow 9/2^+_{gs}$ (top right),
$5/2^+_2 \rightarrow 9/2^+_{gs}$ (bottom left) and
$(7/2,9/2)^+ \rightarrow 9/2^+_{gs}$ (bottom left)
at beam energy $E=3400$ keV.
}
\label{fig:angdist}
\end{figure*}

Typical examples of measured angular distributions are shown in Fig.~\ref{fig:angdist},
showing the $\gamma$--transition angular pattern for the transitions
$5/2^+_1 \rightarrow 9/2^+_{gs}$,
$5/2^+_2 \rightarrow 9/2^+_{gs}$,
$7/2^+_1 \rightarrow 9/2^+_{gs}$, and
$(7/2^+,9/2^+) \rightarrow 9/2^+_{gs}$
at beam energy of $E_p = 3400$ keV. In cases where an angular distribution
was not clearly demonstrated in the data (mainly due to large uncertainties),
an average value was used instead (see e.g. lower right panel in
Fig.~\ref{fig:angdist}). In addition, no $\gamma_0$ was observed in the
spectra, likely due to the large spin difference between the entry state
(1/2$^+$ or 3/2$^+$) and the ground state of \isotope[113]{In} (J$^\pi$=9/2$^+$).
The results for the ground state cross section are tabulated in
Table~\ref{tab:gs_cross_sections} and plotted in Fig.~\ref{fig:scheme_gs}.

\subsection{Activation measurements}

The isomeric transition $1/2^-_1 \rightarrow 9/2^+_{gs}$ is characterized
by a half life of $t_{1/2} = 99.476 (23)$ min. The measurement of the
absolute yield of the particular transition demanded the use of the
activation method. An additional measurement of the cross section
of the isomeric state was performed with the in--beam method that was
discussed in the previous paragraph.

For each beam energy, the target was irradiated for approximately three
half--lives, which is a sufficient irradiation time interval, as after about
$5t_{1/2}$, the process reaches saturation~\cite{iliadis2015nuclear}.
The isomeric cross section was evaluated using the standard relation:
\begin{equation}
\label{eq: activation xs}
\sigma_{is} = \frac{A \lambda e^{\lambda t_w}}{N_t \phi
\epsilon_{abs} I_\gamma (1-e^{-\lambda t_c})(1-e^{- \lambda t_{irr}})}
\end{equation}
where $A$ is the number of events under the corresponding photopeak of the
isomeric transition,
$I_\gamma$ is the probability of $\gamma$--ray emission,
$\lambda$ is the decay constant of the transition,
$N_t$ is the number of target nuclei per unit area,
$\phi$ is the incident proton flux during the irradiation,
$\epsilon_{abs}$ is the absolute efficiency of the detector and
$t_w$, $t_c$, $t_{irr}$ are the waiting (or cooling) time
of the sample, the counting time and the irradiation time of the sample,
respectively. For the present case,
$I_\gamma = 0.6494(17)$ and
$\lambda = 116.133(27) \times 10^{-6}$~s$^{-1}$~\cite{endsf,2010_Blachot}.  

The results for the isomeric cross sections with the activation method
are tabulated in Table~\ref{tab:gs_cross_sections} and plotted in
Fig.~\ref{fig:scheme_activation} (solid diamonds). Errors were evaluated
by considering the uncertainties from photopeak integration, the detector
efficiencies and the charge deposition on the target during the irradiation
of the sample. Cross--section results for the isomeric state deduced from the
in--beam technique taking into account all transitions reaching the isomeric
state are shown in the same figure (empty circles).

\subsection{Total cross--sections and astrophysical \textit{S} factors}

The total cross section of the reaction $^{112}$Cd($p,\gamma$)$^{113}$In,
$\sigma_T$, have been evaluated by adding the cross sections of all transitions
feeding the ground state of the produced nucleus (summing to the in--beam
cross section $\sigma_{gs}$) and the cross--section of the isomeric state,
$\sigma_{is}$, as measured with the activation technique described earlier:
\begin{equation}
\sigma_T = \sigma_{gs} + \sigma_{is}
\end{equation}

The results for the total cross section of the studied reaction are tabulated
in Table~\ref{tab:gs_cross_sections} and plotted in Fig.~\ref{fig:scheme_pg}.
After measuring the total cross section, the astrophysical \textit{S} factor
can be deduced, by means of the relation:
\begin{equation}
S(E)= E \sigma(E)  e^{2\pi \eta}
\label{eq:Sfac}
\end{equation}
where $\eta$ is the Sommerfeld parameter~\cite{2010_Yakovlev}. The results
for the astrophysical \textit{S} factor are also tabulated in
Table~\ref{tab:gs_cross_sections} and plotted in Fig.~\ref{fig:scheme_pgS}.
The particular quantity is important for astrophysical applications, as it
varies smoothly with energy, compared to the cross section, thus allowing
for safer extrapolations to experimentally inaccessible energies, serving
also as a useful quantity for reaction network calculations.

All energies selected for the experiment reside inside the Gamow window
and below the $(p,n)$ reaction threshold at energy of $E\approx 3.4$~MeV~\cite{qvalue}
(see Table~\ref{tab:gs_cross_sections} for details).
\begin{table*}[ht]
\caption{Cross sections, astrophysical \textit{S} factors and isomeric ratios for the studied reaction.}
\begin{ruledtabular}
\begin{tabular}{ccccccccc}
$E_p$ (lab) & $E_{eff}$ (lab) & $E_{eff}$ (c.m.)   & $\sigma_{gs}$ & $\sigma_{is}$ & $\sigma_T$ & \textit{S} factor & $\sigma_{is} / \sigma_{gs}$ & $\sigma_{is} / \sigma_T$ \\
 (MeV) & (MeV) & (MeV) & (mb) & (mb) & (mb) & ($\times 10^8$ MeV b) \\ \hline
 2.800 & 2.771 & 2.746 & $0.0075\pm 0.0005$ & $0.014\pm 0.001$ & $0.021 \pm 0.001$ & $1.60 \pm 0.10$ & $1.8 \pm 0.2$  & $0.64 \pm 0.07$\\
 3.000 & 2.972 & 2.945 & $0.030\pm 0.004$ & $0.050\pm 0.004$ & $0.080 \pm 0.005$ & $2.43 \pm 0.16$ & $ 1.7 \pm 0.2$ & $0.63 \pm 0.07$\\
 3.200 & 3.172 & 3.144 & $0.070\pm 0.004$ & $0.125\pm 0.009$ & $0.195 \pm 0.010$ & $2.59 \pm 0.13$ & $1.8 \pm 0.2$ & $0.64 \pm 0.06$\\
 3.400 & 3.374 & 3.344 & $0.138\pm 0.007$ & $0.265\pm 0.016$  & $0.404 \pm 0.018$ & $2.54 \pm 0.11$ & $1.9 \pm 0.2$ & $0.66 \pm 0.05$
\end{tabular}
\label{tab:gs_cross_sections}
\end{ruledtabular}
\end{table*}

\begin{figure}
\centering
\includegraphics[width=0.48\textwidth]{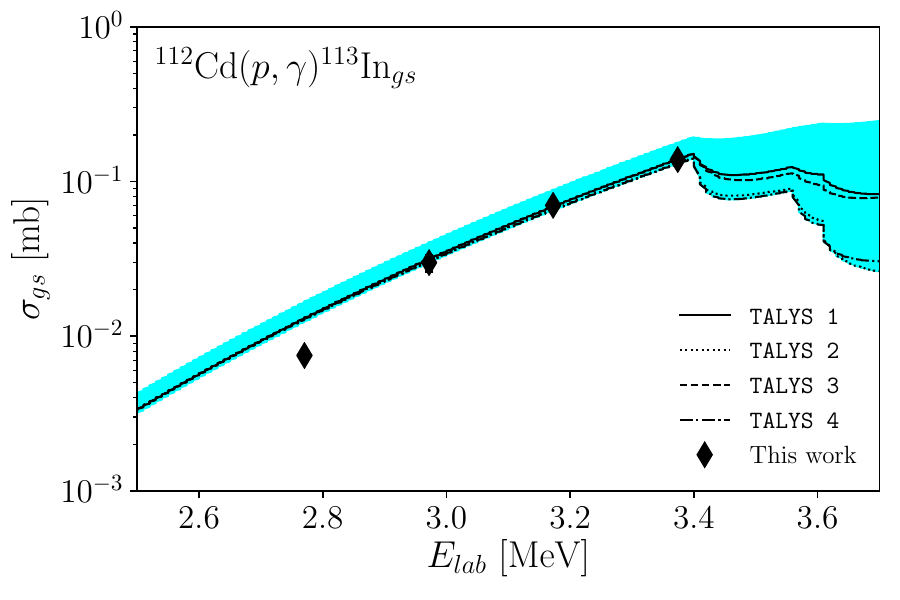}
\caption{Ground--state cross sections for the $(p,\gamma)$ channel deduced
by the in--beam method. Energies are shown in the lab system. The shaded area
corresponds to the full range of calculated values with every combination of models
employed. The lines correspond to the best data--matching calculations, see text
for details.}
\label{fig:scheme_gs}
\end{figure}
\begin{figure}
\centering
\includegraphics[width=0.48\textwidth]{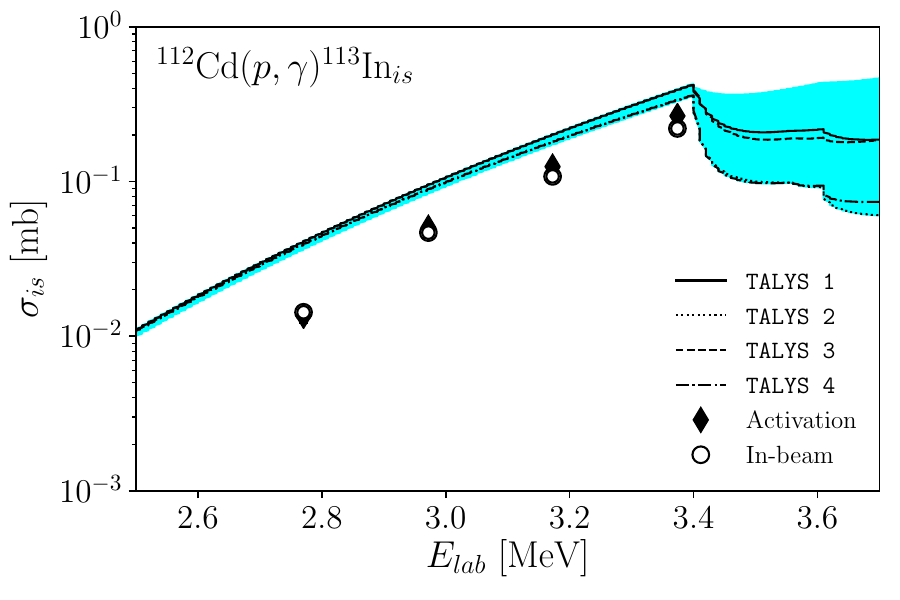}
\caption{Measured isomeric cross sections with both the activation
(solid diamonds) and the in--beam (open circles) methods. The lines and shaded
area are as in Fig.~\ref{fig:scheme_gs}.}
\label{fig:scheme_activation}
\end{figure}

\begin{figure}
\centering
\includegraphics[width=0.48\textwidth]{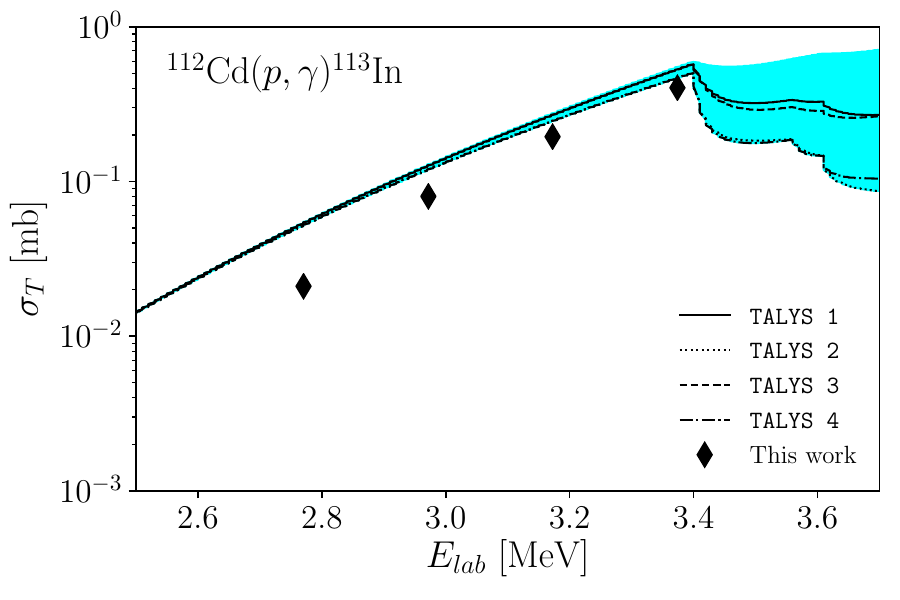}
\caption{As in Fig.~\ref{fig:scheme_gs}, but for the total reaction cross sections
of the $(p,\gamma)$ channel deduced from the in--beam and activation methods.}
\label{fig:scheme_pg}
\end{figure}
\begin{figure}
\centering
\includegraphics[width=0.48\textwidth]{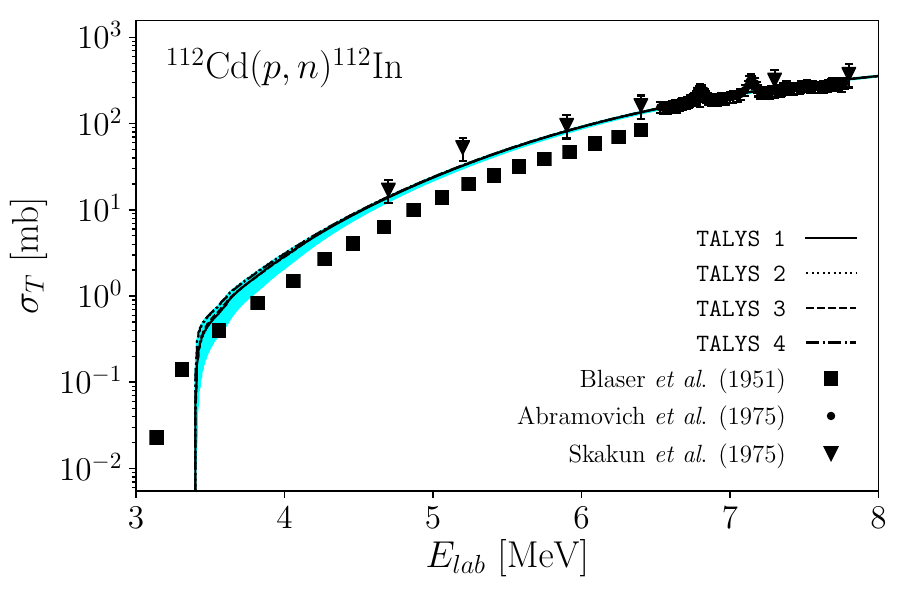}
\caption{Experimental data are compared to \texttt{TALYS} calculations for the total
cross sections of the $(p,n)$ channel. The data have been retrieved from
literature (Blaser \textit{et al.}~\cite{Blaser_1951}, Abramovich
\textit{et al.}~\cite{NSR1975AB09} and Skakun \textit{et al.}~\cite{Skakun_1975}).
See also Fig.~\ref{fig:scheme_gs} for details regarding the shaded area and the
line curves.}
\label{fig:scheme_pn}
\end{figure}

\begin{figure}
\centering
\includegraphics[width=0.48\textwidth]{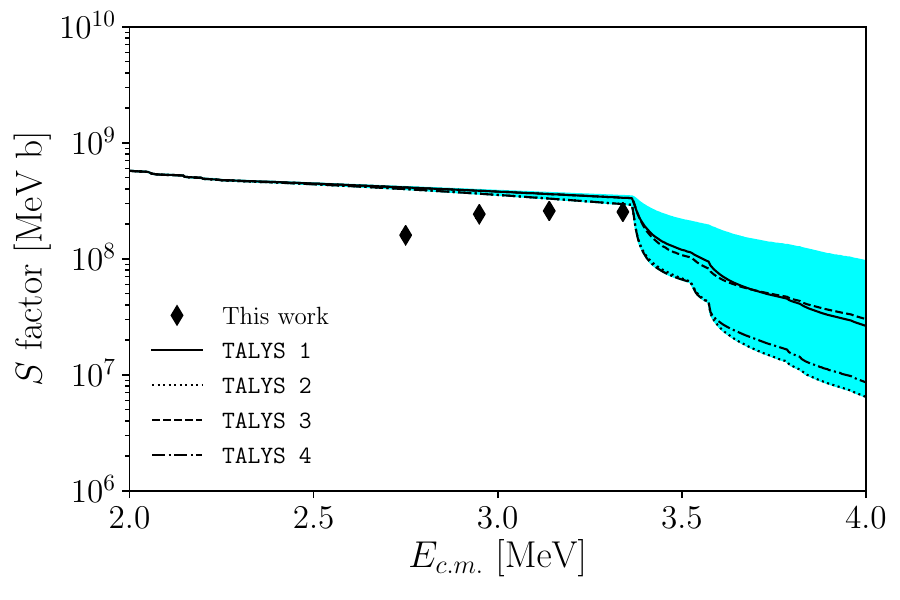}
\caption{As in Fig.~\ref{fig:scheme_pg}, but for astrophysical \textit{S} 
factors. The only difference is that energies are shown in the
center--of--mass system.}
\label{fig:scheme_pgS}
\end{figure}

\begin{figure*}
\centering
\includegraphics[width=0.48\textwidth]{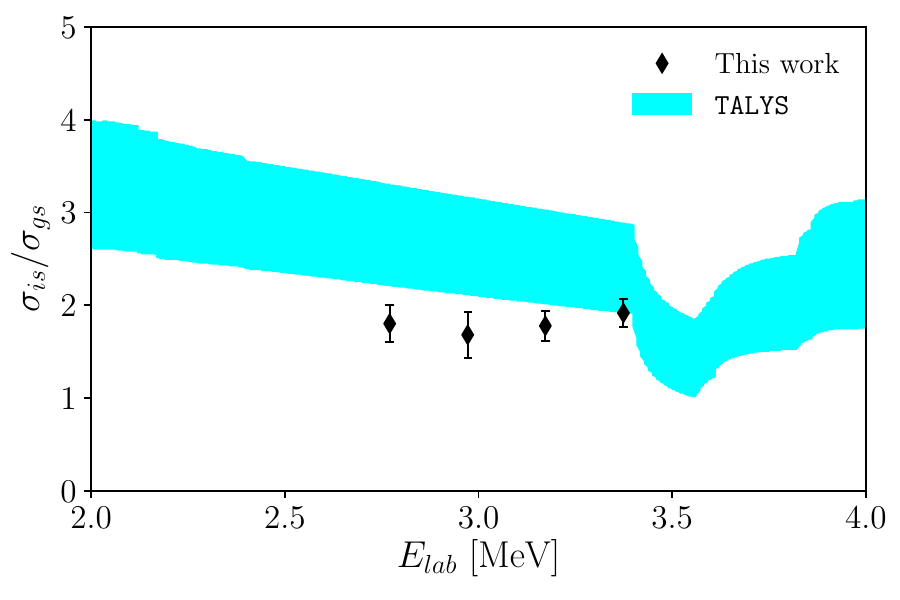}
\includegraphics[width=0.48\textwidth]{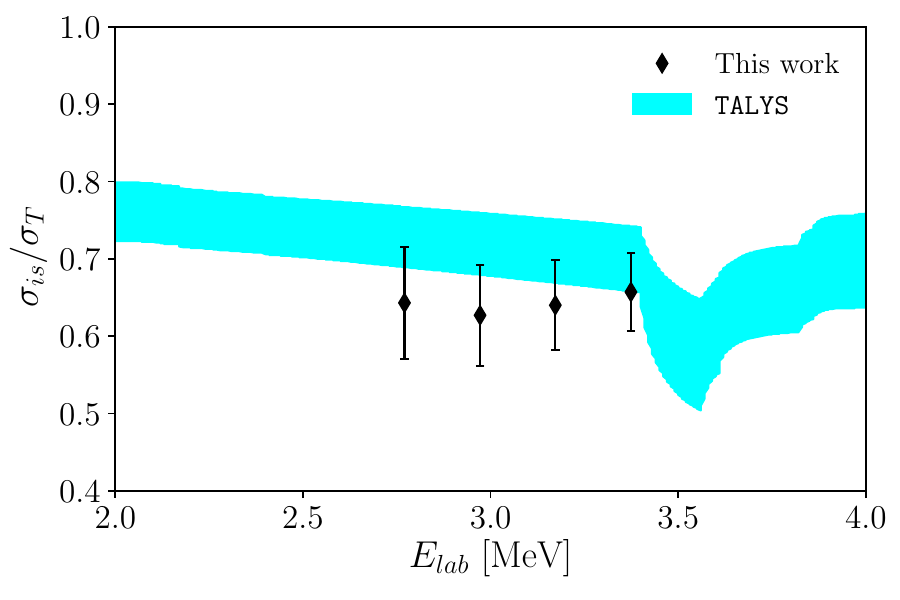}
\caption{Isomeric ratios of the isomeric cross section to the ground state
cross section (left) and of the isomeric cross section to the total cross
section (right) of the reaction $^{112}$Cd($p,\gamma$)$^{113}$In.
Please note the different scales of the $y$ axes.}
\label{fig:isomeric_ratios}
\end{figure*}

\subsection{Hauser--Feshbach calculations with \texttt{TALYS}}

Theoretical calculations using the Hauser--Feshbach statistical model
have been performed with the \texttt{TALYS} v1.9 code~\cite{koning2007talys}.
A total of 96 different combinations of the main ingredients of the model,
i.e. the Optical Potential (OMP) (2 default options),
the Nuclear Level Density (NLD) (6 default options)
and the $\gamma$--ray Strength Function ($\gamma$SF) (8 default options)
have been used. The models used are presented in Table~\ref{tab:talys}.
The calculations were performed using a 5--keV energy step,
between 1.5 and 8.0~MeV using the supercomputing facility \textit{Z--machine}
at NCSR ``Demokritos''.
\begin{table*}
\caption{Models used for the calculations of cross section with
\texttt{TALYS}~\cite{koning2007talys}. In total, results from 96 combinations
are presented in this work.
}
\label{tab:talys}
\begin{tabular}{ccc}
\hline
\hline \Tstrut
\textbf{Optical Model Potential} & \textbf{Nuclear Level Density} & \textbf{$\gamma$ Strength Function} \\ \hline \Tstrut
Koning--Delaroche (KD)~\cite{KONING2003231} & Constant Temperature Model (CTM)~\cite{gilbert1965ctm} & Kopecky--Uhl~\cite{1990Kopecky} \\
Bauge--Delaroche--Girod (BDG)~\cite{2001Bauge} & Back--shifted Fermi gas model (BSFG)~\cite{DILG1973269} & Brink--Axel~\cite{BRINK1957215,AxelPhysRev.126.671}\\
&  Generalized Superfluid Model (GSM)~\cite{1993Ignatyuk} & Hartree--Fock BCS (HFBCS)~\cite{2009Capote}\\
&  Goriely Tables~\cite{GORIELY2001311} & Hartree--Fock--Bogolubov (HFB)~\cite{2009Capote}\\
&  Hilaire Tables~\cite{Goriely_HilairePhysRevC.78.064307} & Goriely Hybrid Model~\cite{1998Goriely}\\
& T--dependent HFB, Gogny force (TDHFB)~\cite{2012Hilaire} & Goriely TDHFB \cite{2012Hilaire}\\
& & T--Dependent RMF~\cite{2008Arteaga}\\
& & Gogny D1M HFB+QRPA~\cite{2014Martini}\\
\hline\hline
\end{tabular}

\end{table*}

Both microscopic and phenomenological models have been used for calculations,
using the default parameters provided by \texttt{TALYS}.
For the OMP,
the phenomenological model of Koning--Delaroche~\cite{KONING2003231},
as well as
the semi--microscopic model of Bauge--Delaroche--Girod~\cite{2001Bauge}
has been used. It is important to note that, at the studied energy range,
which lies below the Coulomb barrier, the OMP, and in particular its
imaginary component, is known to depend strongly on the energy~\cite{arnould2003p}.

All six available NLD models provided by \texttt{TALYS} have been used in the
calculations, namely
the phenomenological Constant--Temperature model (CTM)~\cite{gilbert1965ctm},
the Back--shifted Fermi gas model~\cite{DILG1973269},
the Generalized Superfluid model~\cite{1993Ignatyuk},
the semi--microscopic level density tables of Goriely~\cite{GORIELY2001311},
and Hilaire~\cite{Goriely_HilairePhysRevC.78.064307},
and values using the Time--Dependent Hartree--Fock--Bogolyubov method
combined with the Gogny force~\cite{2012Hilaire}.

Regarding $\gamma$SF models, the Kopecky--Uhl~\cite{1990Kopecky} and
Brink--Axel~\cite{BRINK1957215} generalized lorentzians were used,
as well as values calculated using the Hartree--Fock--BCS and
Hartree--Fock--Bogolyubov methods~\cite{2009Capote}.
Goriely's hybrid model~\cite{1998Goriely}, as well as Goriely's tables
using the temperature--dependent Hartree--Fock--Bogolyubov method
were additionally employed.
Last, models using the Temperature--Dependent Relativistic Mean
Field method~\cite{2012Hilaire} and the Hartree--Fock--Bogolyubov method
along with the Quasi--Random--Phase--Approximation using the Gogny D1M
interaction~\cite{2014Martini} have been considered.

After performing all possible calculations with the models described
above, the maximum and minimum for each energy has been determined,
defining the borders of the light blue area shown in
Figs.~\ref{fig:scheme_gs}--\ref{fig:scheme_pgS}.
The calculations (TALYS~1--4) that best describe the ground--state cross
section, based on direct comparison with the experimental data, have been
also included in the plots:
TALYS~1 and TALYS~2 employ the Koning--Delaroche OMP, while TALYS~3 and
TALYS~4 use the Bauge--Delaroche--Girod OMP;
TALYS~1 and TALYS~3 employ the Generalized Superfluid model NLD and the
HFBCS $\gamma$SF, while TALYS~2 and TALYS~4 use the TDHFB with the Gogny
force NLD model and the Temperature--dependent RMF $\gamma$SF model.

\section{Discussion and Conclusions}
\label{sec:discussion}

In the framework of the present work, an experimental attempt to measure
the total reaction cross section and the \textit{S} factor of the
astrophysically important reaction \isotope[112]{Cd}($p,\gamma$)\isotope[113]{In}
has been carried out for the first time. The cross section was measured
inside the astrophysically relevant energy range, at four beam energies,
namely 2.8, 3.0, 3.2 and 3.4~MeV.

The measurement of the total reaction cross section required the use
of two different techniques. The cross section of all prompt $\gamma$
transitions feeding the ground state of the produced nucleus was determined
using the in--beam $\gamma$--angular distribution method. All visible
transitions in the spectra feeding the isomeric state were included
in the measurement of its cross section. However, due to the significantly
longer half--life of the isomeric state the activation technique was
employed~\cite{1988_Rolfs,2019_Gyurky} additionally and was used to
produce the total cross section. Table~\ref{tab:is_both} lists the
two data sets for each energy value and the \% deviation of the cross
section deduced from the in--beam method from the corresponding value
found with the activation technique.
\begin{table}
\caption{
Isomeric cross sections deduced from the activation and
the in--beam measurements for the four beam energies (lab).
In the far right column, the \% absolute differences of the in--beam
results with respect to the activation results are shown.
The data sets are also shown in Fig.~\ref{fig:scheme_activation}.
}

\begin{ruledtabular}
\begin{tabular}{cccc}
$E_{eff}$ (lab) & $\sigma_{is}$ (Activation) & $\sigma_{is}$ (In--beam) & Deviation \\
(MeV) & (mb) & (mb) & (\%)  \\ \hline
2.771 & 0.014 $\pm$ 0.001 & 0.0143 $\pm$ 0.0008 & 2 \\
2.972 & 0.050 $\pm$ 0.004 &  0.047 $\pm$ 0.004  & 6 \\
3.172 & 0.125 $\pm$ 0.009 &  0.108 $\pm$ 0.006  & 14 \\
3.374 & 0.265 $\pm$ 0.016 &  0.220 $\pm$ 0.011  & 17 \\
\end{tabular}
\end{ruledtabular}
\label{tab:is_both}
\end{table}

The absolute yields of seven (7) transitions feeding directly the ground
state of \isotope[113]{In} have been measured. It has to be stressed that
the cross sections are particularly small (7.5--138~$\upmu$b for the
in--beam measurements; 14--265~$\upmu$b for the activation measurements)
posing a real difficulty in collecting sufficient statistics, especially
for the low--populated states decaying directly to the ground state at
the lowest energy of 2.8~MeV. A few of the corresponding transitions hide
under the background built up in singles mode, thus resulting in some
missing yield. However, in the present work, this missing yield can be
safely considered smaller than the experimental error for the two lower
energies (Fig.~\ref{fig:scheme_activation}).

An alternative experimental approach to remedy all that could possibly
be the application of the $4\pi$ detection method, which simplifies the
tedious data analysis of a complex $\gamma$--ray spectrum, since it results
into a single summing peak. The aforementioned method has been applied
successfully for studies in reactions relevant to the \textit{p}
process~\cite{spyrou2007cross} despite its own constraints, such as the
summing efficiency, which depends on the $\gamma$--decay
scheme~\cite{iliadis2015nuclear}.

As mentioned earlier, the cross section of the isomeric state was measured
using the activation technique in addition to measuring transitions feeding
the isomeric state during the application of the in--beam technique.
Compared to the latter case, in the activation method, there is no
beam--induced background in the spectra and no angular distribution
effect to consider. In the present case, the decay of the \isotope[113]{In}
isomer emits 392--keV $\gamma$ rays, where the efficiencies of the detectors
are relatively better, compared with the higher--energy $\gamma$ transitions
measured with the in--beam method. However, it is of extreme importance
to have accurate knowledge of the half--life and the branching ratios of
the isomeric state, as the measurement explicitly depends on their values
(see Eq.~\ref{eq: activation xs}).

Combining the ground--state cross sections from the in--beam technique
and the isomeric cross sections from the activation technique (see data
listed in Table~\ref{tab:gs_cross_sections}) the total cross sections,
$\sigma_T$, for the reaction \isotope[112]{Cd}($p,\gamma$)\isotope[113]{In}
has been deduced for all four energy values, ranging 21--404 $\upmu$b
(also in Table~\ref{tab:gs_cross_sections}). These results show a smooth
increase with increasing energy as illustrated in Fig.~\ref{fig:scheme_pg}.
The $\sigma_T$ values were used further to calculate the astrophysical
\textit{S} factors by means of Eq.~\ref{eq:Sfac}, also included in
Table~\ref{tab:gs_cross_sections}. The \textit{S} factor values exhibit
an almost constant behavior, except for the lower energy point at beam
energy 2.8 MeV, as it is evident from the data trend in Fig.~\ref{fig:scheme_pgS}.

From the experimental data in Table~\ref{tab:gs_cross_sections}, the
isomeric--to--ground state cross section ratio, $R_{gs}=\sigma_{is}/\sigma_{gs}$,
and the isomeric--to--total cross section ratio, $R_T=\sigma_{is}/\sigma_T$,
can be evaluated, as well. The isomeric cross section ratios are particularly
useful in understanding the transfer of angular momentum in nuclear reactions.
The results are shown in the two far--right columns in the same Table and shown
in Fig.~\ref{fig:isomeric_ratios}. Both ratios remain almost constant at different
energies. Their weighted--averages have been deduced:
$\left(R_{gs}\right)_{avg}=1.82(9)$ and
$\left(R_{T}\right)_{avg} =0.64(3)$.

Theoretical calculations using the Hauser--Feshbach model have been
performed, incorporating all possible combinations of the default \texttt{TALYS}
parameters of the models tabulated in Table~\ref{tab:talys}. The range of
all calculations for each energy for the total cross section is plotted
in Fig.~\ref{fig:scheme_pg}, along with the experimental data. As expected,
below the energy threshold of the ($p,n$) channel ($E_{thresh} = 3397.39$ keV),
the dependence from the NLD and $\gamma$SF models is relatively weak.
In this energy range, the cross section depends almost exclusively on
the choice of the OMP parameters, as it is evident in the convergence
of all calculations at low energies.

Despite some overestimation, the theoretical predictions describe the trend
of the experimental data fairly well (Figs.~\ref{fig:scheme_gs}--\ref{fig:scheme_pg}).
TALYS~1--4 calculations agree well with the in--beam results with some small
overestimation at 2.8~MeV for the ground state (Fig.~\ref{fig:scheme_gs}).
For the isomeric state, the theoretical trend is in fair agreement with
the experimental results except the lowest energy point
(Fig.~\ref{fig:scheme_activation}), despite an
overall overestimation of the cross section data, which is subsequently
reflected on the total cross section (Fig.~\ref{fig:scheme_pg}). There is
no obvious reason for this minor disagreement from an experimental point
of view. To further investigate the situation, the employed \texttt{TALYS}
models have to be revisited more carefully, especially in regards of
the OMP involved. Such disagreements have been observed in other cases in
this mass regime (see e.g.~\cite{2001_Gyurky}, the review article by
Gy\"urky~\textit{et al.} \cite{2019_Gyurky} and references therein) and
require careful consideration of the statistical uncertainties included
in the models, as well as more detailed experimental work.

Along these lines, the ($p,n$) channel can offer some useful insights.
Calculations for the cross sections of the ($p,n$) channel have been
performed simultaneously with the ($p,\gamma$) channel. These calculations
are compared with existing experimental data, as shown in Fig.~\ref{fig:scheme_pn}.
The theoretical results seem to agree well with the data above 6.0~MeV,
but theoretical calculations seem to diverge from the data below that
energy value down to the ($p,n$) energy threshold. Also, two different
sets of experimental data, those by
Blaser \emph{et al.}~\cite{Blaser_1951} and
Skakun \emph{et al.}~\cite{Skakun_1975},
seem to significantly disagree with one another in the energy range between
4.5 and 6.2~MeV, and both with the present calculation (more the former,
less the latter). However, the combinations TALYS~1--4, which best describe
the ground--state cross section of the ($p,\gamma$) channel, seem to also
describe the data of Skakun~\emph{et al.} rather well.

It could be argued that the observed disagreement between the data and
the theoretical calculations is due to the fact that the incorporated
phenomenological and semi--microscopic OMPs have been optimized at
significantly higher energy range than the one the present study focuses on.
Consequently, an extrapolation to energies lower than the ($p,n$) threshold
may be responsible for the overestimation of the experimentally deduced
total reaction cross section data. However, it has to be noted that a
full sensitivity analysis of the OMP parameters is beyond the scope of
this work, as this would require careful consideration of all models
involved in the calculation, scrutinizing the respective statistical
uncertainties, and potentially fine--tuning the numerous model parameters.

Overall, the present work provides the first set of experimentally
deduced cross sections, astrophysical \textit{S} factors and isomeric
ratios in \isotope[113]{In} populated in a proton--capture radiative
reaction. The new information can support the improvement of reaction
network calculations around the \textit{p} nucleus $\isotope[113]{In}$.
Certainly, further investigation is required in this region of the
nuclear chart, both theoretically and experimentally, to provide firm
insight at the driving mechanisms behind the \textit{p} process reaction
network, as well as to improve the phenomenological parts of the Optical
Model Potentials in an energy region where a scarcity of experimental
data, even for stable nuclei, still persists. 

\begin{acknowledgments}
The authors gratefully acknowledge the technical and scientific 
staff of the Tandem Accelerator Laboratory at NCSR ``Demokritos''
for their support during the experiment and useful discussions.
We thank
E. Mavrommatis for useful discussions, 
C.~Markou and K.~Pikounis for assistance in using the supercomputing
facility at NCSR ``D'' and Dr. K.~Mergia for providing access to the
XRF spectroscopy station.
A.~Khaliel acknowledges support from the Hellenic Foundation for Research and
Innovation~(HFRI) and the General Secretariat for Research and Technology
(GSRT), under the PhD Fellowship grant (GA. no. 74117/2017), and is thankful
to the organizers, lecturers and fellow trainees of the ChETEC training school
``An experiment of Nuclear Physics for Astrophysics using direct methods'',
hosted by IFIN-HH of Bucharest-Magurele, for the fruitful discussions on
the activation method.
P.~Tsavalas has performed work within the framework of the EUROfusion Consortium
which has received funding from the Euratom research and training programme
2014-2018 and 2019-2020 under grant agreement No 633053. The views and
opinions expressed herein do not necessarily reflect those of the European
Commission.
We are grateful to an anonymous reviewer for providing constructive comments
resulting in an overall improvement of the present work.
\end{acknowledgments}

%\bibliographystyle{normal}
%\bibliography{113In_paper}

\bibliographystyle{apsrev}
\bibliography{bibliography}

\begin{thebibliography}{78}
\expandafter\ifx\csname natexlab\endcsname\relax\def\natexlab#1{#1}\fi
\expandafter\ifx\csname bibnamefont\endcsname\relax
  \def\bibnamefont#1{#1}\fi
\expandafter\ifx\csname bibfnamefont\endcsname\relax
  \def\bibfnamefont#1{#1}\fi
\expandafter\ifx\csname citenamefont\endcsname\relax
  \def\citenamefont#1{#1}\fi
\expandafter\ifx\csname url\endcsname\relax
  \def\url#1{\texttt{#1}}\fi
\expandafter\ifx\csname urlprefix\endcsname\relax\def\urlprefix{URL }\fi
\providecommand{\bibinfo}[2]{#2}
\providecommand{\eprint}[2][]{\url{#2}}

\bibitem[{\citenamefont{Burbidge et~al.}(1957)\citenamefont{Burbidge, Burbidge,
  Fowler, and Hoyle}}]{burbidge1957synthesis}
\bibinfo{author}{\bibfnamefont{E.~M.} \bibnamefont{Burbidge}},
  \bibinfo{author}{\bibfnamefont{G.~R.} \bibnamefont{Burbidge}},
  \bibinfo{author}{\bibfnamefont{W.~A.} \bibnamefont{Fowler}},
  \bibnamefont{and} \bibinfo{author}{\bibfnamefont{F.}~\bibnamefont{Hoyle}},
  \bibinfo{journal}{{Rev. Mod. Phys.}} \textbf{\bibinfo{volume}{29}},
  \bibinfo{pages}{547} (\bibinfo{year}{1957}).

\bibitem[{\citenamefont{Cameron}(1957)}]{cameron1957nuclear}
\bibinfo{author}{\bibfnamefont{A.~G.~W.} \bibnamefont{Cameron}},
  \bibinfo{journal}{{Publ. Astron. Soc. Pac.}} \textbf{\bibinfo{volume}{69}},
  \bibinfo{pages}{201} (\bibinfo{year}{1957}).

\bibitem[{\citenamefont{Lodders et~al.}(2009)\citenamefont{Lodders, Palme, and
  Gail}}]{lodders20094}
\bibinfo{author}{\bibfnamefont{K.}~\bibnamefont{Lodders}},
  \bibinfo{author}{\bibfnamefont{H.}~\bibnamefont{Palme}}, \bibnamefont{and}
  \bibinfo{author}{\bibfnamefont{H.-P.} \bibnamefont{Gail}}, in
  \emph{\bibinfo{booktitle}{Solar system}} (\bibinfo{publisher}{Springer},
  \bibinfo{year}{2009}), pp. \bibinfo{pages}{712--770}.

\bibitem[{\citenamefont{Arnould and Goriely}(2003)}]{arnould2003p}
\bibinfo{author}{\bibfnamefont{M.}~\bibnamefont{Arnould}} \bibnamefont{and}
  \bibinfo{author}{\bibfnamefont{S.}~\bibnamefont{Goriely}},
  \bibinfo{journal}{{Phys. Rep.}} \textbf{\bibinfo{volume}{384}},
  \bibinfo{pages}{1} (\bibinfo{year}{2003}).

\bibitem[{\citenamefont{Rauscher et~al.}(2013)\citenamefont{Rauscher, Dauphas,
  Dillmann, Fr{\"o}hlich, F{\"u}l{\"o}p, and
  Gy{\"u}rky}}]{rauscher2013constraining}
\bibinfo{author}{\bibfnamefont{T.}~\bibnamefont{Rauscher}},
  \bibinfo{author}{\bibfnamefont{N.}~\bibnamefont{Dauphas}},
  \bibinfo{author}{\bibfnamefont{I.}~\bibnamefont{Dillmann}},
  \bibinfo{author}{\bibfnamefont{C.}~\bibnamefont{Fr{\"o}hlich}},
  \bibinfo{author}{\bibfnamefont{Z.}~\bibnamefont{F{\"u}l{\"o}p}},
  \bibnamefont{and}
  \bibinfo{author}{\bibfnamefont{G.}~\bibnamefont{Gy{\"u}rky}},
  \bibinfo{journal}{{Rep. Prog. Phys.}} \textbf{\bibinfo{volume}{76}},
  \bibinfo{pages}{066201} (\bibinfo{year}{2013}).

\bibitem[{\citenamefont{Woosley and Howard}(1978)}]{woosley1978p}
\bibinfo{author}{\bibfnamefont{S.}~\bibnamefont{Woosley}} \bibnamefont{and}
  \bibinfo{author}{\bibfnamefont{W.}~\bibnamefont{Howard}},
  \bibinfo{journal}{{Astrophys. J. Suppl. Ser.}} \textbf{\bibinfo{volume}{36}},
  \bibinfo{pages}{285} (\bibinfo{year}{1978}).

\bibitem[{\citenamefont{Travaglio et~al.}(2011)\citenamefont{Travaglio,
  R{\"o}pke, Gallino, and Hillebrandt}}]{travaglio2011type}
\bibinfo{author}{\bibfnamefont{C.}~\bibnamefont{Travaglio}},
  \bibinfo{author}{\bibfnamefont{F.~K.} \bibnamefont{R{\"o}pke}},
  \bibinfo{author}{\bibfnamefont{R.}~\bibnamefont{Gallino}}, \bibnamefont{and}
  \bibinfo{author}{\bibfnamefont{W.}~\bibnamefont{Hillebrandt}},
  \bibinfo{journal}{{Astrophys. J.}} \textbf{\bibinfo{volume}{739}},
  \bibinfo{pages}{93} (\bibinfo{year}{2011}).

\bibitem[{\citenamefont{Schatz et~al.}(1999)\citenamefont{Schatz, Bildsten,
  Cumming, and Wiescher}}]{schatz1999rapid}
\bibinfo{author}{\bibfnamefont{H.}~\bibnamefont{Schatz}},
  \bibinfo{author}{\bibfnamefont{L.}~\bibnamefont{Bildsten}},
  \bibinfo{author}{\bibfnamefont{A.}~\bibnamefont{Cumming}}, \bibnamefont{and}
  \bibinfo{author}{\bibfnamefont{M.}~\bibnamefont{Wiescher}},
  \bibinfo{journal}{{Astrophys. J.}} \textbf{\bibinfo{volume}{524}},
  \bibinfo{pages}{1014} (\bibinfo{year}{1999}).

\bibitem[{\citenamefont{Goriely et~al.}(2002)\citenamefont{Goriely, Jos{\'e},
  Hernanz, Rayet, and Arnould}}]{goriely2002he}
\bibinfo{author}{\bibfnamefont{S.}~\bibnamefont{Goriely}},
  \bibinfo{author}{\bibfnamefont{J.}~\bibnamefont{Jos{\'e}}},
  \bibinfo{author}{\bibfnamefont{M.}~\bibnamefont{Hernanz}},
  \bibinfo{author}{\bibfnamefont{M.}~\bibnamefont{Rayet}}, \bibnamefont{and}
  \bibinfo{author}{\bibfnamefont{M.}~\bibnamefont{Arnould}},
  \bibinfo{journal}{{Astron. Astrophys.}} \textbf{\bibinfo{volume}{383}},
  \bibinfo{pages}{L27} (\bibinfo{year}{2002}).

\bibitem[{\citenamefont{Fr{\"o}hlich et~al.}(2006)\citenamefont{Fr{\"o}hlich,
  Martinez-Pinedo, Liebend{\"o}rfer, Thielemann, Bravo, Hix, Langanke, and
  Zinner}}]{frohlich2006neutrino}
\bibinfo{author}{\bibfnamefont{C.}~\bibnamefont{Fr{\"o}hlich}},
  \bibinfo{author}{\bibfnamefont{G.}~\bibnamefont{Martinez-Pinedo}},
  \bibinfo{author}{\bibfnamefont{M.}~\bibnamefont{Liebend{\"o}rfer}},
  \bibinfo{author}{\bibfnamefont{F.-K.} \bibnamefont{Thielemann}},
  \bibinfo{author}{\bibfnamefont{E.}~\bibnamefont{Bravo}},
  \bibinfo{author}{\bibfnamefont{W.~R.} \bibnamefont{Hix}},
  \bibinfo{author}{\bibfnamefont{K.}~\bibnamefont{Langanke}}, \bibnamefont{and}
  \bibinfo{author}{\bibfnamefont{N.~T.} \bibnamefont{Zinner}},
  \bibinfo{journal}{{Phys. Rev. Lett.}} \textbf{\bibinfo{volume}{96}},
  \bibinfo{pages}{142502} (\bibinfo{year}{2006}).

\bibitem[{\citenamefont{Pruet et~al.}(2006)\citenamefont{Pruet, Hoffman,
  Woosley, Janka, and Buras}}]{pruet2006nucleosynthesis}
\bibinfo{author}{\bibfnamefont{J.}~\bibnamefont{Pruet}},
  \bibinfo{author}{\bibfnamefont{R.~D.} \bibnamefont{Hoffman}},
  \bibinfo{author}{\bibfnamefont{S.~E.} \bibnamefont{Woosley}},
  \bibinfo{author}{\bibfnamefont{H.-T.} \bibnamefont{Janka}}, \bibnamefont{and}
  \bibinfo{author}{\bibfnamefont{R.}~\bibnamefont{Buras}},
  \bibinfo{journal}{{Astrophys. J.}} \textbf{\bibinfo{volume}{644}},
  \bibinfo{pages}{1028} (\bibinfo{year}{2006}).

\bibitem[{\citenamefont{Wanajo}(2006)}]{wanajo2006rp}
\bibinfo{author}{\bibfnamefont{S.}~\bibnamefont{Wanajo}},
  \bibinfo{journal}{{Astrophys. J.}} \textbf{\bibinfo{volume}{647}},
  \bibinfo{pages}{1323} (\bibinfo{year}{2006}).

\bibitem[{\citenamefont{Rapp et~al.}(2006)\citenamefont{Rapp, G{\"o}rres,
  Wiescher, Schatz, and K{\"a}ppeler}}]{rapp2006sensitivity}
\bibinfo{author}{\bibfnamefont{W.}~\bibnamefont{Rapp}},
  \bibinfo{author}{\bibfnamefont{J.}~\bibnamefont{G{\"o}rres}},
  \bibinfo{author}{\bibfnamefont{M.}~\bibnamefont{Wiescher}},
  \bibinfo{author}{\bibfnamefont{H.}~\bibnamefont{Schatz}}, \bibnamefont{and}
  \bibinfo{author}{\bibfnamefont{F.}~\bibnamefont{K{\"a}ppeler}},
  \bibinfo{journal}{{Astrophys. J.}} \textbf{\bibinfo{volume}{653}},
  \bibinfo{pages}{474} (\bibinfo{year}{2006}).

\bibitem[{\citenamefont{Hauser and Feshbach}(1952)}]{hauser1952inelastic}
\bibinfo{author}{\bibfnamefont{W.}~\bibnamefont{Hauser}} \bibnamefont{and}
  \bibinfo{author}{\bibfnamefont{H.}~\bibnamefont{Feshbach}},
  \bibinfo{journal}{{Phys. Rev.}} \textbf{\bibinfo{volume}{87}},
  \bibinfo{pages}{366} (\bibinfo{year}{1952}).

\bibitem[{\citenamefont{Jos{\'e} and Iliadis}(2011)}]{jose2011nuclear}
\bibinfo{author}{\bibfnamefont{J.}~\bibnamefont{Jos{\'e}}} \bibnamefont{and}
  \bibinfo{author}{\bibfnamefont{C.}~\bibnamefont{Iliadis}},
  \bibinfo{journal}{{Rep. Prog. Phys.}} \textbf{\bibinfo{volume}{74}},
  \bibinfo{pages}{096901} (\bibinfo{year}{2011}).

\bibitem[{\citenamefont{Ward and Beer}(1981)}]{ward1981origin}
\bibinfo{author}{\bibfnamefont{R.~A.} \bibnamefont{Ward}} \bibnamefont{and}
  \bibinfo{author}{\bibfnamefont{H.}~\bibnamefont{Beer}},
  \bibinfo{journal}{{Astron. Astrophys.}} \textbf{\bibinfo{volume}{103}},
  \bibinfo{pages}{189} (\bibinfo{year}{1981}).

\bibitem[{\citenamefont{Woosley et~al.}(1990)\citenamefont{Woosley, Hartmann,
  Hoffman, and Haxton}}]{woosley1990nu}
\bibinfo{author}{\bibfnamefont{S.~E.} \bibnamefont{Woosley}},
  \bibinfo{author}{\bibfnamefont{D.~H.} \bibnamefont{Hartmann}},
  \bibinfo{author}{\bibfnamefont{R.~D.} \bibnamefont{Hoffman}},
  \bibnamefont{and} \bibinfo{author}{\bibfnamefont{W.~C.}
  \bibnamefont{Haxton}}, \bibinfo{journal}{{Astrophys. J.}}
  \textbf{\bibinfo{volume}{356}}, \bibinfo{pages}{272} (\bibinfo{year}{1990}).

\bibitem[{\citenamefont{N{\'e}meth et~al.}(1994)\citenamefont{N{\'e}meth,
  K{\"a}ppeler, Theis, Belgya, and Yates}}]{nemeth1994nucleosynthesis}
\bibinfo{author}{\bibfnamefont{Z.}~\bibnamefont{N{\'e}meth}},
  \bibinfo{author}{\bibfnamefont{F.}~\bibnamefont{K{\"a}ppeler}},
  \bibinfo{author}{\bibfnamefont{C.}~\bibnamefont{Theis}},
  \bibinfo{author}{\bibfnamefont{T.}~\bibnamefont{Belgya}}, \bibnamefont{and}
  \bibinfo{author}{\bibfnamefont{S.~W.} \bibnamefont{Yates}},
  \bibinfo{journal}{{Astrophys. J.}} \textbf{\bibinfo{volume}{426}},
  \bibinfo{pages}{357} (\bibinfo{year}{1994}).

\bibitem[{\citenamefont{Theis et~al.}(1998)\citenamefont{Theis, K{\"a}ppeler,
  Wisshak, and Voss}}]{theis1998puzzling}
\bibinfo{author}{\bibfnamefont{C.}~\bibnamefont{Theis}},
  \bibinfo{author}{\bibfnamefont{F.}~\bibnamefont{K{\"a}ppeler}},
  \bibinfo{author}{\bibfnamefont{K.}~\bibnamefont{Wisshak}}, \bibnamefont{and}
  \bibinfo{author}{\bibfnamefont{F.}~\bibnamefont{Voss}},
  \bibinfo{journal}{{Astrophys. J.}} \textbf{\bibinfo{volume}{500}},
  \bibinfo{pages}{1039} (\bibinfo{year}{1998}).

\bibitem[{\citenamefont{Dillmann et~al.}(2006)\citenamefont{Dillmann, Heil,
  K{\"a}ppeler, Plag, Rauscher, and Thielemann}}]{dillmann2006kadonis}
\bibinfo{author}{\bibfnamefont{I.}~\bibnamefont{Dillmann}},
  \bibinfo{author}{\bibfnamefont{M.}~\bibnamefont{Heil}},
  \bibinfo{author}{\bibfnamefont{F.}~\bibnamefont{K{\"a}ppeler}},
  \bibinfo{author}{\bibfnamefont{R.}~\bibnamefont{Plag}},
  \bibinfo{author}{\bibfnamefont{T.}~\bibnamefont{Rauscher}}, \bibnamefont{and}
  \bibinfo{author}{\bibfnamefont{F.-K.} \bibnamefont{Thielemann}}, in
  \emph{\bibinfo{booktitle}{{AIP Conf. Proc.}}} (\bibinfo{organization}{AIP},
  \bibinfo{year}{2006}), vol. \bibinfo{volume}{819}, pp.
  \bibinfo{pages}{123--127}.

\bibitem[{\citenamefont{Dillmann et~al.}(2007)\citenamefont{Dillmann, Rauscher,
  Heil, K{\"a}ppeler, Rapp, and Thielemann}}]{dillmann2007p}
\bibinfo{author}{\bibfnamefont{I.}~\bibnamefont{Dillmann}},
  \bibinfo{author}{\bibfnamefont{T.}~\bibnamefont{Rauscher}},
  \bibinfo{author}{\bibfnamefont{M.}~\bibnamefont{Heil}},
  \bibinfo{author}{\bibfnamefont{F.}~\bibnamefont{K{\"a}ppeler}},
  \bibinfo{author}{\bibfnamefont{W.}~\bibnamefont{Rapp}}, \bibnamefont{and}
  \bibinfo{author}{\bibfnamefont{F.~K.} \bibnamefont{Thielemann}},
  \bibinfo{journal}{{J. Phys. G}} \textbf{\bibinfo{volume}{35}},
  \bibinfo{pages}{014029} (\bibinfo{year}{2007}).

\bibitem[{\citenamefont{Dillmann et~al.}(2008)\citenamefont{Dillmann,
  K{\"a}ppeler, Rauscher, Thielemann, Gallino, and
  Bisterzo}}]{dillmann2008there}
\bibinfo{author}{\bibfnamefont{I.}~\bibnamefont{Dillmann}},
  \bibinfo{author}{\bibfnamefont{F.}~\bibnamefont{K{\"a}ppeler}},
  \bibinfo{author}{\bibfnamefont{T.}~\bibnamefont{Rauscher}},
  \bibinfo{author}{\bibfnamefont{F.~K.} \bibnamefont{Thielemann}},
  \bibinfo{author}{\bibfnamefont{R.}~\bibnamefont{Gallino}}, \bibnamefont{and}
  \bibinfo{author}{\bibfnamefont{S.}~\bibnamefont{Bisterzo}}, in
  \emph{\bibinfo{booktitle}{{PoS(NIC X)091}}} (\bibinfo{year}{2008}),
  \bibinfo{note}{doi: 10.22323/1.053.0091}.

\bibitem[{\citenamefont{Wanajo et~al.}(2011)\citenamefont{Wanajo, Janka, and
  Kubono}}]{wanajo2011uncertainties}
\bibinfo{author}{\bibfnamefont{S.}~\bibnamefont{Wanajo}},
  \bibinfo{author}{\bibfnamefont{H.-T.} \bibnamefont{Janka}}, \bibnamefont{and}
  \bibinfo{author}{\bibfnamefont{S.}~\bibnamefont{Kubono}},
  \bibinfo{journal}{{Astrophys. J.}} \textbf{\bibinfo{volume}{729}},
  \bibinfo{pages}{46} (\bibinfo{year}{2011}).

\bibitem[{\citenamefont{Fujimoto et~al.}(2007)\citenamefont{Fujimoto,
  Hashimoto, Kotake, and Yamada}}]{fujimoto2007heavy}
\bibinfo{author}{\bibfnamefont{S.}~\bibnamefont{Fujimoto}},
  \bibinfo{author}{\bibfnamefont{M.}~\bibnamefont{Hashimoto}},
  \bibinfo{author}{\bibfnamefont{K.}~\bibnamefont{Kotake}}, \bibnamefont{and}
  \bibinfo{author}{\bibfnamefont{S.}~\bibnamefont{Yamada}},
  \bibinfo{journal}{{Astrophys. J.}} \textbf{\bibinfo{volume}{656}},
  \bibinfo{pages}{382} (\bibinfo{year}{2007}).

\bibitem[{\citenamefont{Woosley}(1993)}]{woosley1993gamma}
\bibinfo{author}{\bibfnamefont{S.~E.} \bibnamefont{Woosley}},
  \bibinfo{journal}{{Astrophys. J.}} \textbf{\bibinfo{volume}{405}},
  \bibinfo{pages}{273} (\bibinfo{year}{1993}).

\bibitem[{\citenamefont{Babishov and Kopytin}(2006)}]{babishov2006model}
\bibinfo{author}{\bibfnamefont{E.~M.} \bibnamefont{Babishov}} \bibnamefont{and}
  \bibinfo{author}{\bibfnamefont{I.~V.} \bibnamefont{Kopytin}},
  \bibinfo{journal}{{Astron. Rep.}} \textbf{\bibinfo{volume}{50}},
  \bibinfo{pages}{569} (\bibinfo{year}{2006}).

\bibitem[{\citenamefont{Kopytin and Hussain}(2013)}]{kopytin2013role}
\bibinfo{author}{\bibfnamefont{I.~V.} \bibnamefont{Kopytin}} \bibnamefont{and}
  \bibinfo{author}{\bibfnamefont{I.~A.} \bibnamefont{Hussain}},
  \bibinfo{journal}{{Phys. Atom. Nucl.}} \textbf{\bibinfo{volume}{76}},
  \bibinfo{pages}{476} (\bibinfo{year}{2013}).

\bibitem[{\citenamefont{Pignatari et~al.}(2016)\citenamefont{Pignatari,
  G{\"o}bel, Reifarth, and Travaglio}}]{pignatari2016production}
\bibinfo{author}{\bibfnamefont{M.}~\bibnamefont{Pignatari}},
  \bibinfo{author}{\bibfnamefont{K.}~\bibnamefont{G{\"o}bel}},
  \bibinfo{author}{\bibfnamefont{R.}~\bibnamefont{Reifarth}}, \bibnamefont{and}
  \bibinfo{author}{\bibfnamefont{C.}~\bibnamefont{Travaglio}},
  \bibinfo{journal}{{Int. J. Mod. Phys. E}} \textbf{\bibinfo{volume}{25}},
  \bibinfo{pages}{1630003} (\bibinfo{year}{2016}).

\bibitem[{\citenamefont{Harissopulos et~al.}(2016)\citenamefont{Harissopulos,
  Spyrou, Foteinou, Axiotis, Provatas, and Demetriou}}]{harisso113In}
\bibinfo{author}{\bibfnamefont{S.}~\bibnamefont{Harissopulos}},
  \bibinfo{author}{\bibfnamefont{A.}~\bibnamefont{Spyrou}},
  \bibinfo{author}{\bibfnamefont{V.}~\bibnamefont{Foteinou}},
  \bibinfo{author}{\bibfnamefont{M.}~\bibnamefont{Axiotis}},
  \bibinfo{author}{\bibfnamefont{G.}~\bibnamefont{Provatas}}, \bibnamefont{and}
  \bibinfo{author}{\bibfnamefont{P.}~\bibnamefont{Demetriou}},
  \bibinfo{journal}{{Phys. Rev. C}} \textbf{\bibinfo{volume}{93}},
  \bibinfo{pages}{025804} (\bibinfo{year}{2016}).

\bibitem[{\citenamefont{Kiss et~al.}(2013)\citenamefont{Kiss, Mohr,
  F{\"u}l{\"o}p, Rauscher, Gy{\"u}rky, Sz{\"u}cs, Hal{\'a}sz, Somorjai,
  Ornelas, Yal{\c{c}}{\i}n et~al.}}]{kiss2013high}
\bibinfo{author}{\bibfnamefont{G.~G.} \bibnamefont{Kiss}},
  \bibinfo{author}{\bibfnamefont{P.}~\bibnamefont{Mohr}},
  \bibinfo{author}{\bibfnamefont{Z.}~\bibnamefont{F{\"u}l{\"o}p}},
  \bibinfo{author}{\bibfnamefont{T.}~\bibnamefont{Rauscher}},
  \bibinfo{author}{\bibfnamefont{G.}~\bibnamefont{Gy{\"u}rky}},
  \bibinfo{author}{\bibfnamefont{T.}~\bibnamefont{Sz{\"u}cs}},
  \bibinfo{author}{\bibfnamefont{Z.}~\bibnamefont{Hal{\'a}sz}},
  \bibinfo{author}{\bibfnamefont{E.}~\bibnamefont{Somorjai}},
  \bibinfo{author}{\bibfnamefont{A.}~\bibnamefont{Ornelas}},
  \bibinfo{author}{\bibfnamefont{C.}~\bibnamefont{Yal{\c{c}}{\i}n}},
  \bibnamefont{et~al.}, \bibinfo{journal}{{Phys. Rev. C}}
  \textbf{\bibinfo{volume}{88}}, \bibinfo{pages}{045804}
  (\bibinfo{year}{2013}).

\bibitem[{\citenamefont{Yal{\c{c}}{\i}n
  et~al.}(2009)\citenamefont{Yal{\c{c}}{\i}n, G{\"u}ray, {\"O}zkan, Kutlu,
  Gy{\"u}rky, Farkas, Kiss, F{\"u}l{\"o}p, Simon, Somorjai
  et~al.}}]{yalccin2009odd}
\bibinfo{author}{\bibfnamefont{C.}~\bibnamefont{Yal{\c{c}}{\i}n}},
  \bibinfo{author}{\bibfnamefont{R.}~\bibnamefont{G{\"u}ray}},
  \bibinfo{author}{\bibfnamefont{N.}~\bibnamefont{{\"O}zkan}},
  \bibinfo{author}{\bibfnamefont{S.}~\bibnamefont{Kutlu}},
  \bibinfo{author}{\bibfnamefont{G.}~\bibnamefont{Gy{\"u}rky}},
  \bibinfo{author}{\bibfnamefont{J.}~\bibnamefont{Farkas}},
  \bibinfo{author}{\bibfnamefont{G.}~\bibnamefont{Kiss}},
  \bibinfo{author}{\bibfnamefont{Z.}~\bibnamefont{F{\"u}l{\"o}p}},
  \bibinfo{author}{\bibfnamefont{A.}~\bibnamefont{Simon}},
  \bibinfo{author}{\bibfnamefont{E.}~\bibnamefont{Somorjai}},
  \bibnamefont{et~al.}, \bibinfo{journal}{{Phys. Rev. C}}
  \textbf{\bibinfo{volume}{79}}, \bibinfo{pages}{065801}
  (\bibinfo{year}{2009}).

\bibitem[{\citenamefont{Shan et~al.}(2018)\citenamefont{Shan, Musthafa,
  Najmunnisa, Aslam, Rajesh, Hajara, Surendran, Nair, Shanbagh, and
  Ghugre}}]{2018_MuhammedShan}
\bibinfo{author}{\bibfnamefont{P.~T.~M.} \bibnamefont{Shan}},
  \bibinfo{author}{\bibfnamefont{M.~M.} \bibnamefont{Musthafa}},
  \bibinfo{author}{\bibfnamefont{T.}~\bibnamefont{Najmunnisa}},
  \bibinfo{author}{\bibfnamefont{P.~M.} \bibnamefont{Aslam}},
  \bibinfo{author}{\bibfnamefont{K.~K.} \bibnamefont{Rajesh}},
  \bibinfo{author}{\bibfnamefont{K.}~\bibnamefont{Hajara}},
  \bibinfo{author}{\bibfnamefont{P.}~\bibnamefont{Surendran}},
  \bibinfo{author}{\bibfnamefont{J.~P.} \bibnamefont{Nair}},
  \bibinfo{author}{\bibfnamefont{A.}~\bibnamefont{Shanbagh}}, \bibnamefont{and}
  \bibinfo{author}{\bibfnamefont{S.}~\bibnamefont{Ghugre}},
  \bibinfo{journal}{{Nucl. Phys. A}}  (\bibinfo{year}{2018}), ISSN
  \bibinfo{issn}{0375-9474}.

\bibitem[{\citenamefont{{Blaser} et~al.}(1951)\citenamefont{{Blaser}, {Boehm},
  {Marmier}, and {Peaslee}}}]{Blaser_1951}
\bibinfo{author}{\bibfnamefont{J.~P.} \bibnamefont{{Blaser}}},
  \bibinfo{author}{\bibfnamefont{F.}~\bibnamefont{{Boehm}}},
  \bibinfo{author}{\bibfnamefont{P.}~\bibnamefont{{Marmier}}},
  \bibnamefont{and} \bibinfo{author}{\bibfnamefont{D.~C.}
  \bibnamefont{{Peaslee}}}, \bibinfo{journal}{{Helv. Phys. Acta}}
  \textbf{\bibinfo{volume}{24}}, \bibinfo{pages}{3} (\bibinfo{year}{1951}).

\bibitem[{\citenamefont{{Abramovich} et~al.}(1975)\citenamefont{{Abramovich},
  {Guzhovskii}, {Zvenigorodskii}, and {Trusillo}}}]{NSR1975AB09}
\bibinfo{author}{\bibfnamefont{S.~N.} \bibnamefont{{Abramovich}}},
  \bibinfo{author}{\bibfnamefont{B.~Y.} \bibnamefont{{Guzhovskii}}},
  \bibinfo{author}{\bibfnamefont{A.~G.} \bibnamefont{{Zvenigorodskii}}},
  \bibnamefont{and} \bibinfo{author}{\bibfnamefont{S.~V.}
  \bibnamefont{{Trusillo}}}, \bibinfo{journal}{{Izv. Akad. Nauk}}
  \textbf{\bibinfo{volume}{SSSR}}, \bibinfo{pages}{Ser.Fiz. 39, 1688}
  (\bibinfo{year}{1975}).

\bibitem[{\citenamefont{{Skakun} et~al.}(1975)\citenamefont{{Skakun},
  {Klyucharev}, {Rakivnenko}, and {Romanii}}}]{Skakun_1975}
\bibinfo{author}{\bibfnamefont{E.~A.} \bibnamefont{{Skakun}}},
  \bibinfo{author}{\bibfnamefont{A.~P.} \bibnamefont{{Klyucharev}}},
  \bibinfo{author}{\bibfnamefont{Y.~N.} \bibnamefont{{Rakivnenko}}},
  \bibnamefont{and} \bibinfo{author}{\bibfnamefont{I.~A.}
  \bibnamefont{{Romanii}}}, \bibinfo{journal}{{Izv. Akad. Nauk}}
  \textbf{\bibinfo{volume}{SSSR}}, \bibinfo{pages}{Ser.Fiz. 39, 24}
  (\bibinfo{year}{1975}).

\bibitem[{\citenamefont{Nedorezov et~al.}(2019)\citenamefont{Nedorezov,
  Konobeevski, Polonski, Ponomarev, Savel’ev, Solodukhov, Tsymbalov, Turinge,
  Zuyev, and Gorlova}}]{2019_Nedorezov}
\bibinfo{author}{\bibfnamefont{V.}~\bibnamefont{Nedorezov}},
  \bibinfo{author}{\bibfnamefont{E.}~\bibnamefont{Konobeevski}},
  \bibinfo{author}{\bibfnamefont{A.}~\bibnamefont{Polonski}},
  \bibinfo{author}{\bibfnamefont{V.}~\bibnamefont{Ponomarev}},
  \bibinfo{author}{\bibfnamefont{A.}~\bibnamefont{Savel’ev}},
  \bibinfo{author}{\bibfnamefont{G.}~\bibnamefont{Solodukhov}},
  \bibinfo{author}{\bibfnamefont{I.}~\bibnamefont{Tsymbalov}},
  \bibinfo{author}{\bibfnamefont{A.}~\bibnamefont{Turinge}},
  \bibinfo{author}{\bibfnamefont{S.}~\bibnamefont{Zuyev}}, \bibnamefont{and}
  \bibinfo{author}{\bibfnamefont{D.}~\bibnamefont{Gorlova}},
  \bibinfo{journal}{{Phys. Scripta}} \textbf{\bibinfo{volume}{94}},
  \bibinfo{pages}{015303} (\bibinfo{year}{2019}).

\bibitem[{\citenamefont{Rauscher}(2006)}]{2006_Rauscher}
\bibinfo{author}{\bibfnamefont{T.}~\bibnamefont{Rauscher}},
  \bibinfo{journal}{Phys. Rev. C} \textbf{\bibinfo{volume}{73}},
  \bibinfo{pages}{015804} (\bibinfo{year}{2006}).

\bibitem[{\citenamefont{Hayakawa et~al.}(2016)\citenamefont{Hayakawa, Toh,
  Huang, Shizuma, Kimura, Nakamura, Harada, Iwamoto, Chiba, and
  Kajino}}]{hayakawa2016measurement}
\bibinfo{author}{\bibfnamefont{T.}~\bibnamefont{Hayakawa}},
  \bibinfo{author}{\bibfnamefont{Y.}~\bibnamefont{Toh}},
  \bibinfo{author}{\bibfnamefont{M.}~\bibnamefont{Huang}},
  \bibinfo{author}{\bibfnamefont{T.}~\bibnamefont{Shizuma}},
  \bibinfo{author}{\bibfnamefont{A.}~\bibnamefont{Kimura}},
  \bibinfo{author}{\bibfnamefont{S.}~\bibnamefont{Nakamura}},
  \bibinfo{author}{\bibfnamefont{H.}~\bibnamefont{Harada}},
  \bibinfo{author}{\bibfnamefont{N.}~\bibnamefont{Iwamoto}},
  \bibinfo{author}{\bibfnamefont{S.}~\bibnamefont{Chiba}}, \bibnamefont{and}
  \bibinfo{author}{\bibfnamefont{T.}~\bibnamefont{Kajino}},
  \bibinfo{journal}{{Phys. Rev. C}} \textbf{\bibinfo{volume}{94}},
  \bibinfo{pages}{055803} (\bibinfo{year}{2016}).

\bibitem[{qva()}]{qvalue}
\emph{\bibinfo{title}{{NNDC} {O}nline {D}ata {S}ervice, \emph{Q}-value
  calculator}}, \urlprefix\url{http://www.nndc.bnl.gov/qcalc/}.

\bibitem[{\citenamefont{{Gy\"urky, Gy.} et~al.}(2019)\citenamefont{{Gy\"urky,
  Gy.}, {F\"ul\"op, Zs.}, {K\"appeler, F.}, {Kiss, G. G.}, and {Wallner,
  A.}}}]{2019_Gyurky}
\bibinfo{author}{\bibnamefont{{Gy\"urky, Gy.}}},
  \bibinfo{author}{\bibnamefont{{F\"ul\"op, Zs.}}},
  \bibinfo{author}{\bibnamefont{{K\"appeler, F.}}},
  \bibinfo{author}{\bibnamefont{{Kiss, G. G.}}}, \bibnamefont{and}
  \bibinfo{author}{\bibnamefont{{Wallner, A.}}}, \bibinfo{journal}{Eur. Phys.
  J. A} \textbf{\bibinfo{volume}{55}}, \bibinfo{pages}{41}
  (\bibinfo{year}{2019}).

\bibitem[{\citenamefont{Khaliel et~al.}(2017)\citenamefont{Khaliel,
  Mertzimekis, Asimakopoulou, Kanellakopoulos, Lagaki, Psaltis, Psyrra, and
  Mavrommatis}}]{khaliel2017first}
\bibinfo{author}{\bibfnamefont{A.}~\bibnamefont{Khaliel}},
  \bibinfo{author}{\bibfnamefont{T.~J.} \bibnamefont{Mertzimekis}},
  \bibinfo{author}{\bibfnamefont{E.-M.} \bibnamefont{Asimakopoulou}},
  \bibinfo{author}{\bibfnamefont{A.}~\bibnamefont{Kanellakopoulos}},
  \bibinfo{author}{\bibfnamefont{V.}~\bibnamefont{Lagaki}},
  \bibinfo{author}{\bibfnamefont{A.}~\bibnamefont{Psaltis}},
  \bibinfo{author}{\bibfnamefont{I.}~\bibnamefont{Psyrra}}, \bibnamefont{and}
  \bibinfo{author}{\bibfnamefont{E.}~\bibnamefont{Mavrommatis}},
  \bibinfo{journal}{{Phys. Rev. C}} \textbf{\bibinfo{volume}{96}},
  \bibinfo{pages}{035806} (\bibinfo{year}{2017}).

\bibitem[{\citenamefont{Ziegler et~al.}(2010)\citenamefont{Ziegler, Ziegler,
  and Biersack}}]{ZIEGLER20101818}
\bibinfo{author}{\bibfnamefont{J.~F.} \bibnamefont{Ziegler}},
  \bibinfo{author}{\bibfnamefont{M.~D.} \bibnamefont{Ziegler}},
  \bibnamefont{and} \bibinfo{author}{\bibfnamefont{J.~P.}
  \bibnamefont{Biersack}}, \bibinfo{journal}{Nuclear Instruments and Methods in
  Physics Research Section B: Beam Interactions with Materials and Atoms}
  \textbf{\bibinfo{volume}{268}}, \bibinfo{pages}{1818 }
  (\bibinfo{year}{2010}), ISSN \bibinfo{issn}{0168-583X}, \bibinfo{note}{19th
  International Conference on Ion Beam Analysis}.

\bibitem[{\citenamefont{Galanopoulos et~al.}(2003)\citenamefont{Galanopoulos,
  Demetriou, Kokkoris, Harissopulos, Kunz, Fey, Hammer, Gy{\"u}rky,
  F{\"u}l{\"o}p, Somorjai et~al.}}]{galanopoulos200388}
\bibinfo{author}{\bibfnamefont{S.}~\bibnamefont{Galanopoulos}},
  \bibinfo{author}{\bibfnamefont{P.}~\bibnamefont{Demetriou}},
  \bibinfo{author}{\bibfnamefont{M.}~\bibnamefont{Kokkoris}},
  \bibinfo{author}{\bibfnamefont{S.}~\bibnamefont{Harissopulos}},
  \bibinfo{author}{\bibfnamefont{R.}~\bibnamefont{Kunz}},
  \bibinfo{author}{\bibfnamefont{M.}~\bibnamefont{Fey}},
  \bibinfo{author}{\bibfnamefont{J.~W.} \bibnamefont{Hammer}},
  \bibinfo{author}{\bibfnamefont{G.}~\bibnamefont{Gy{\"u}rky}},
  \bibinfo{author}{\bibfnamefont{Z.}~\bibnamefont{F{\"u}l{\"o}p}},
  \bibinfo{author}{\bibfnamefont{E.}~\bibnamefont{Somorjai}},
  \bibnamefont{et~al.}, \bibinfo{journal}{{Phys. Rev. C}}
  \textbf{\bibinfo{volume}{67}}, \bibinfo{pages}{015801}
  (\bibinfo{year}{2003}).

\bibitem[{\citenamefont{Sauerwein et~al.}(2012)\citenamefont{Sauerwein, Endres,
  Netterdon, Zilges, Foteinou, Provatas, Konstantinopoulos, Axiotis, Ashley,
  Harissopulos et~al.}}]{sauerwein2012investigation}
\bibinfo{author}{\bibfnamefont{A.}~\bibnamefont{Sauerwein}},
  \bibinfo{author}{\bibfnamefont{J.}~\bibnamefont{Endres}},
  \bibinfo{author}{\bibfnamefont{L.}~\bibnamefont{Netterdon}},
  \bibinfo{author}{\bibfnamefont{A.}~\bibnamefont{Zilges}},
  \bibinfo{author}{\bibfnamefont{V.}~\bibnamefont{Foteinou}},
  \bibinfo{author}{\bibfnamefont{G.}~\bibnamefont{Provatas}},
  \bibinfo{author}{\bibfnamefont{T.}~\bibnamefont{Konstantinopoulos}},
  \bibinfo{author}{\bibfnamefont{M.}~\bibnamefont{Axiotis}},
  \bibinfo{author}{\bibfnamefont{S.~F.} \bibnamefont{Ashley}},
  \bibinfo{author}{\bibfnamefont{S.}~\bibnamefont{Harissopulos}},
  \bibnamefont{et~al.}, \bibinfo{journal}{{Phys. Rev. C}}
  \textbf{\bibinfo{volume}{86}}, \bibinfo{pages}{035802}
  (\bibinfo{year}{2012}).

\bibitem[{nnd()}]{nndc}
\emph{\bibinfo{title}{National {N}uclear {D}ata {C}enter}},
  \urlprefix\url{http://www.nndc.bnl.gov/nudat2}.

\bibitem[{\citenamefont{Yal\c{c}in et~al.}(2009)\citenamefont{Yal\c{c}in,
  G\"uray, \"Ozkan, Kutlu, Gy\"urky, Farkas, Kiss, F\"ul\"op, Simon, Somorjai
  et~al.}}]{Yalcin2009}
\bibinfo{author}{\bibfnamefont{C.}~\bibnamefont{Yal\c{c}in}},
  \bibinfo{author}{\bibfnamefont{R.~T.} \bibnamefont{G\"uray}},
  \bibinfo{author}{\bibfnamefont{N.}~\bibnamefont{\"Ozkan}},
  \bibinfo{author}{\bibfnamefont{S.}~\bibnamefont{Kutlu}},
  \bibinfo{author}{\bibfnamefont{G.}~\bibnamefont{Gy\"urky}},
  \bibinfo{author}{\bibfnamefont{J.}~\bibnamefont{Farkas}},
  \bibinfo{author}{\bibfnamefont{G.~G.} \bibnamefont{Kiss}},
  \bibinfo{author}{\bibfnamefont{Z.}~\bibnamefont{F\"ul\"op}},
  \bibinfo{author}{\bibfnamefont{A.}~\bibnamefont{Simon}},
  \bibinfo{author}{\bibfnamefont{E.}~\bibnamefont{Somorjai}},
  \bibnamefont{et~al.}, \bibinfo{journal}{Phys. Rev. C}
  \textbf{\bibinfo{volume}{79}}, \bibinfo{pages}{065801}
  (\bibinfo{year}{2009}).

\bibitem[{\citenamefont{Kiss et~al.}(2011)\citenamefont{Kiss, Rauscher,
  Sz\"ucs, Kert\'esz, F\"ul\"op, Gy\"urky, Fr\"ohlich, Farkas, Elekes, and
  Somorjai}}]{Kiss2011}
\bibinfo{author}{\bibfnamefont{G.~G.} \bibnamefont{Kiss}},
  \bibinfo{author}{\bibfnamefont{T.}~\bibnamefont{Rauscher}},
  \bibinfo{author}{\bibfnamefont{T.}~\bibnamefont{Sz\"ucs}},
  \bibinfo{author}{\bibfnamefont{Z.}~\bibnamefont{Kert\'esz}},
  \bibinfo{author}{\bibfnamefont{Z.}~\bibnamefont{F\"ul\"op}},
  \bibinfo{author}{\bibfnamefont{G.}~\bibnamefont{Gy\"urky}},
  \bibinfo{author}{\bibfnamefont{C.}~\bibnamefont{Fr\"ohlich}},
  \bibinfo{author}{\bibfnamefont{J.}~\bibnamefont{Farkas}},
  \bibinfo{author}{\bibfnamefont{Z.}~\bibnamefont{Elekes}}, \bibnamefont{and}
  \bibinfo{author}{\bibfnamefont{E.}~\bibnamefont{Somorjai}},
  \bibinfo{journal}{{Phys. Lett. B}} \textbf{\bibinfo{volume}{695}},
  \bibinfo{pages}{419 } (\bibinfo{year}{2011}), ISSN \bibinfo{issn}{0370-2693}.

\bibitem[{\citenamefont{Dillmann et~al.}(2011)\citenamefont{Dillmann, Coquard,
  Domingo-Pardo, K\"appeler, Marganiec, Uberseder, Giesen, Heiske, Feinberg,
  Hentschel et~al.}}]{Dillman2011}
\bibinfo{author}{\bibfnamefont{I.}~\bibnamefont{Dillmann}},
  \bibinfo{author}{\bibfnamefont{L.}~\bibnamefont{Coquard}},
  \bibinfo{author}{\bibfnamefont{C.}~\bibnamefont{Domingo-Pardo}},
  \bibinfo{author}{\bibfnamefont{F.}~\bibnamefont{K\"appeler}},
  \bibinfo{author}{\bibfnamefont{J.}~\bibnamefont{Marganiec}},
  \bibinfo{author}{\bibfnamefont{E.}~\bibnamefont{Uberseder}},
  \bibinfo{author}{\bibfnamefont{U.}~\bibnamefont{Giesen}},
  \bibinfo{author}{\bibfnamefont{A.}~\bibnamefont{Heiske}},
  \bibinfo{author}{\bibfnamefont{G.}~\bibnamefont{Feinberg}},
  \bibinfo{author}{\bibfnamefont{D.}~\bibnamefont{Hentschel}},
  \bibnamefont{et~al.}, \bibinfo{journal}{{Phys. Rev. C}}
  \textbf{\bibinfo{volume}{84}}, \bibinfo{pages}{015802}
  (\bibinfo{year}{2011}).

\bibitem[{\citenamefont{Sauerwein et~al.}(2011)\citenamefont{Sauerwein, Becker,
  Dombrowski, Elvers, Endres, Giesen, Hasper, Hennig, Netterdon, Rauscher
  et~al.}}]{Sauerwein2011}
\bibinfo{author}{\bibfnamefont{A.}~\bibnamefont{Sauerwein}},
  \bibinfo{author}{\bibfnamefont{H.-W.} \bibnamefont{Becker}},
  \bibinfo{author}{\bibfnamefont{H.}~\bibnamefont{Dombrowski}},
  \bibinfo{author}{\bibfnamefont{M.}~\bibnamefont{Elvers}},
  \bibinfo{author}{\bibfnamefont{J.}~\bibnamefont{Endres}},
  \bibinfo{author}{\bibfnamefont{U.}~\bibnamefont{Giesen}},
  \bibinfo{author}{\bibfnamefont{J.}~\bibnamefont{Hasper}},
  \bibinfo{author}{\bibfnamefont{A.}~\bibnamefont{Hennig}},
  \bibinfo{author}{\bibfnamefont{L.}~\bibnamefont{Netterdon}},
  \bibinfo{author}{\bibfnamefont{T.}~\bibnamefont{Rauscher}},
  \bibnamefont{et~al.}, \bibinfo{journal}{{Phys. Rev. C}}
  \textbf{\bibinfo{volume}{84}}, \bibinfo{pages}{045808}
  (\bibinfo{year}{2011}).

\bibitem[{\citenamefont{Hal\'asz et~al.}(2012)\citenamefont{Hal\'asz, Gy\"urky,
  Farkas, F\"ul\"op, Sz\"ucs, Somorjai, and Rauscher}}]{Halasz2012}
\bibinfo{author}{\bibfnamefont{Z.}~\bibnamefont{Hal\'asz}},
  \bibinfo{author}{\bibfnamefont{G.}~\bibnamefont{Gy\"urky}},
  \bibinfo{author}{\bibfnamefont{J.}~\bibnamefont{Farkas}},
  \bibinfo{author}{\bibfnamefont{Z.}~\bibnamefont{F\"ul\"op}},
  \bibinfo{author}{\bibfnamefont{T.}~\bibnamefont{Sz\"ucs}},
  \bibinfo{author}{\bibfnamefont{E.}~\bibnamefont{Somorjai}}, \bibnamefont{and}
  \bibinfo{author}{\bibfnamefont{T.}~\bibnamefont{Rauscher}},
  \bibinfo{journal}{{Phys. Rev. C}} \textbf{\bibinfo{volume}{85}},
  \bibinfo{pages}{025804} (\bibinfo{year}{2012}).

\bibitem[{\citenamefont{Netterdon et~al.}(2013)\citenamefont{Netterdon,
  Demetriou, Endres, Giesen, Kiss, Sauerwein, Sz\"ucs, Zell, and
  Zilges}}]{Netterdon2013}
\bibinfo{author}{\bibfnamefont{L.}~\bibnamefont{Netterdon}},
  \bibinfo{author}{\bibfnamefont{P.}~\bibnamefont{Demetriou}},
  \bibinfo{author}{\bibfnamefont{J.}~\bibnamefont{Endres}},
  \bibinfo{author}{\bibfnamefont{U.}~\bibnamefont{Giesen}},
  \bibinfo{author}{\bibfnamefont{G.}~\bibnamefont{Kiss}},
  \bibinfo{author}{\bibfnamefont{A.}~\bibnamefont{Sauerwein}},
  \bibinfo{author}{\bibfnamefont{T.}~\bibnamefont{Sz\"ucs}},
  \bibinfo{author}{\bibfnamefont{K.}~\bibnamefont{Zell}}, \bibnamefont{and}
  \bibinfo{author}{\bibfnamefont{A.}~\bibnamefont{Zilges}},
  \bibinfo{journal}{{Nucl. Phys. A}} \textbf{\bibinfo{volume}{916}},
  \bibinfo{pages}{149} (\bibinfo{year}{2013}), ISSN \bibinfo{issn}{0375--9474}.

\bibitem[{\citenamefont{Netterdon et~al.}(2014)\citenamefont{Netterdon, Endres,
  Kiss, Mayer, Rauscher, Scholz, Sonnabend, T\"or\"ok, and
  Zilges}}]{Netterdon2014}
\bibinfo{author}{\bibfnamefont{L.}~\bibnamefont{Netterdon}},
  \bibinfo{author}{\bibfnamefont{A.}~\bibnamefont{Endres}},
  \bibinfo{author}{\bibfnamefont{G.~G.} \bibnamefont{Kiss}},
  \bibinfo{author}{\bibfnamefont{J.}~\bibnamefont{Mayer}},
  \bibinfo{author}{\bibfnamefont{T.}~\bibnamefont{Rauscher}},
  \bibinfo{author}{\bibfnamefont{P.}~\bibnamefont{Scholz}},
  \bibinfo{author}{\bibfnamefont{K.}~\bibnamefont{Sonnabend}},
  \bibinfo{author}{\bibfnamefont{Z.}~\bibnamefont{T\"or\"ok}},
  \bibnamefont{and} \bibinfo{author}{\bibfnamefont{A.}~\bibnamefont{Zilges}},
  \bibinfo{journal}{{Phys. Rev. C}} \textbf{\bibinfo{volume}{90}},
  \bibinfo{pages}{035806} (\bibinfo{year}{2014}).

\bibitem[{\citenamefont{G\"uray et~al.}(2015)\citenamefont{G\"uray, \"Ozkan,
  Yal\c{c}in, Rauscher, Gy\"urky, Farkas, F\"ul\"op, Hal\'asz, and
  Somorjai}}]{Guray2015}
\bibinfo{author}{\bibfnamefont{R.~T.} \bibnamefont{G\"uray}},
  \bibinfo{author}{\bibfnamefont{N.}~\bibnamefont{\"Ozkan}},
  \bibinfo{author}{\bibfnamefont{C.}~\bibnamefont{Yal\c{c}in}},
  \bibinfo{author}{\bibfnamefont{T.}~\bibnamefont{Rauscher}},
  \bibinfo{author}{\bibfnamefont{G.}~\bibnamefont{Gy\"urky}},
  \bibinfo{author}{\bibfnamefont{J.}~\bibnamefont{Farkas}},
  \bibinfo{author}{\bibfnamefont{Z.}~\bibnamefont{F\"ul\"op}},
  \bibinfo{author}{\bibfnamefont{Z.}~\bibnamefont{Hal\'asz}}, \bibnamefont{and}
  \bibinfo{author}{\bibfnamefont{E.}~\bibnamefont{Somorjai}},
  \bibinfo{journal}{{Phys. Rev. C}} \textbf{\bibinfo{volume}{91}},
  \bibinfo{pages}{055809} (\bibinfo{year}{2015}).

\bibitem[{\citenamefont{Kinoshita et~al.}(2016)\citenamefont{Kinoshita,
  Hayashi, Ueno, Yatsu, Yokoyama, and Takahashi}}]{Kinoshita2016}
\bibinfo{author}{\bibfnamefont{N.}~\bibnamefont{Kinoshita}},
  \bibinfo{author}{\bibfnamefont{K.}~\bibnamefont{Hayashi}},
  \bibinfo{author}{\bibfnamefont{S.}~\bibnamefont{Ueno}},
  \bibinfo{author}{\bibfnamefont{Y.}~\bibnamefont{Yatsu}},
  \bibinfo{author}{\bibfnamefont{A.}~\bibnamefont{Yokoyama}}, \bibnamefont{and}
  \bibinfo{author}{\bibfnamefont{N.}~\bibnamefont{Takahashi}},
  \bibinfo{journal}{{Phys. Rev. C}} \textbf{\bibinfo{volume}{93}},
  \bibinfo{pages}{025801} (\bibinfo{year}{2016}).

\bibitem[{\citenamefont{Gy{\"u}rky et~al.}(2003)\citenamefont{Gy{\"u}rky,
  F{\"u}l{\"o}p, Somorjai, Kokkoris, Galanopoulos, Demetriou, Harissopulos,
  Rauscher, and Goriely}}]{gyurky2003proton}
\bibinfo{author}{\bibfnamefont{G.}~\bibnamefont{Gy{\"u}rky}},
  \bibinfo{author}{\bibfnamefont{Z.}~\bibnamefont{F{\"u}l{\"o}p}},
  \bibinfo{author}{\bibfnamefont{E.}~\bibnamefont{Somorjai}},
  \bibinfo{author}{\bibfnamefont{M.}~\bibnamefont{Kokkoris}},
  \bibinfo{author}{\bibfnamefont{S.}~\bibnamefont{Galanopoulos}},
  \bibinfo{author}{\bibfnamefont{P.}~\bibnamefont{Demetriou}},
  \bibinfo{author}{\bibfnamefont{S.}~\bibnamefont{Harissopulos}},
  \bibinfo{author}{\bibfnamefont{T.}~\bibnamefont{Rauscher}}, \bibnamefont{and}
  \bibinfo{author}{\bibfnamefont{S.}~\bibnamefont{Goriely}},
  \bibinfo{journal}{{Phys. Rev. C}} \textbf{\bibinfo{volume}{68}},
  \bibinfo{pages}{055803} (\bibinfo{year}{2003}).

\bibitem[{\citenamefont{Rolfs and Rodney}(1988)}]{1988_Rolfs}
\bibinfo{author}{\bibfnamefont{C.}~\bibnamefont{Rolfs}} \bibnamefont{and}
  \bibinfo{author}{\bibfnamefont{W.}~\bibnamefont{Rodney}},
  \emph{\bibinfo{title}{Cauldrons in the Cosmos. Chicago}}
  (\bibinfo{publisher}{University of Chicago Press}, \bibinfo{year}{1988}),
  \bibinfo{note}{iSBN 0-226-72456-5}.

\bibitem[{\citenamefont{Iliadis}(2015)}]{iliadis2015nuclear}
\bibinfo{author}{\bibfnamefont{C.}~\bibnamefont{Iliadis}},
  \emph{\bibinfo{title}{Nuclear physics of stars}} (\bibinfo{publisher}{John
  Wiley \& Sons}, \bibinfo{year}{2015}).

\bibitem[{end()}]{endsf}
\bibinfo{howpublished}{The Evaluated Nuclear Structure Data File (ENSDF),
  http://www.nndc.bnl.gov/ensdf/},
  \urlprefix\url{http://www.nndc.bnl.gov/ensdf/}.

\bibitem[{\citenamefont{Blachot}(2010)}]{2010_Blachot}
\bibinfo{author}{\bibfnamefont{J.}~\bibnamefont{Blachot}},
  \bibinfo{journal}{{Nucl. Data Sheets}} \textbf{\bibinfo{volume}{111}},
  \bibinfo{pages}{1471} (\bibinfo{year}{2010}).

\bibitem[{\citenamefont{Yakovlev et~al.}(2010)\citenamefont{Yakovlev, Beard,
  Gasques, and Wiescher}}]{2010_Yakovlev}
\bibinfo{author}{\bibfnamefont{D.~G.} \bibnamefont{Yakovlev}},
  \bibinfo{author}{\bibfnamefont{M.}~\bibnamefont{Beard}},
  \bibinfo{author}{\bibfnamefont{L.~R.} \bibnamefont{Gasques}},
  \bibnamefont{and} \bibinfo{author}{\bibfnamefont{M.}~\bibnamefont{Wiescher}},
  \bibinfo{journal}{{Phys. Rev. C}} \textbf{\bibinfo{volume}{82}},
  \bibinfo{pages}{044609} (\bibinfo{year}{2010}).

\bibitem[{\citenamefont{Koning et~al.}(2007)\citenamefont{Koning, Hilaire, and
  Duijvestijn}}]{koning2007talys}
\bibinfo{author}{\bibfnamefont{A.~J.} \bibnamefont{Koning}},
  \bibinfo{author}{\bibfnamefont{S.}~\bibnamefont{Hilaire}}, \bibnamefont{and}
  \bibinfo{author}{\bibfnamefont{M.~C.} \bibnamefont{Duijvestijn}}, in
  \emph{\bibinfo{booktitle}{International Conference on Nuclear Data for
  Science and Technology}} (\bibinfo{organization}{EDP Sciences},
  \bibinfo{year}{2007}), pp. \bibinfo{pages}{211--214}.

\bibitem[{\citenamefont{Koning and Delaroche}(2003)}]{KONING2003231}
\bibinfo{author}{\bibfnamefont{A.~J.} \bibnamefont{Koning}} \bibnamefont{and}
  \bibinfo{author}{\bibfnamefont{J.~P.} \bibnamefont{Delaroche}},
  \bibinfo{journal}{{Nucl. Phys. A}} \textbf{\bibinfo{volume}{713}},
  \bibinfo{pages}{231 } (\bibinfo{year}{2003}), ISSN \bibinfo{issn}{0375-9474}.

\bibitem[{\citenamefont{Gilbert and Cameron}(1965)}]{gilbert1965ctm}
\bibinfo{author}{\bibfnamefont{A.}~\bibnamefont{Gilbert}} \bibnamefont{and}
  \bibinfo{author}{\bibfnamefont{A.~G.~W.} \bibnamefont{Cameron}},
  \bibinfo{journal}{{Can. J. Phys.}} \textbf{\bibinfo{volume}{43}},
  \bibinfo{pages}{1446} (\bibinfo{year}{1965}).

\bibitem[{\citenamefont{Kopecky and Uhl}(1990)}]{1990Kopecky}
\bibinfo{author}{\bibfnamefont{J.}~\bibnamefont{Kopecky}} \bibnamefont{and}
  \bibinfo{author}{\bibfnamefont{M.}~\bibnamefont{Uhl}},
  \bibinfo{journal}{{Phys. Rev. C}} \textbf{\bibinfo{volume}{41}},
  \bibinfo{pages}{1941} (\bibinfo{year}{1990}).

\bibitem[{\citenamefont{Bauge et~al.}(2001)\citenamefont{Bauge, Delaroche, and
  Girod}}]{2001Bauge}
\bibinfo{author}{\bibfnamefont{E.}~\bibnamefont{Bauge}},
  \bibinfo{author}{\bibfnamefont{J.~P.} \bibnamefont{Delaroche}},
  \bibnamefont{and} \bibinfo{author}{\bibfnamefont{M.}~\bibnamefont{Girod}},
  \bibinfo{journal}{{Phys. Rev. C}} \textbf{\bibinfo{volume}{63}},
  \bibinfo{pages}{024607} (\bibinfo{year}{2001}).

\bibitem[{\citenamefont{Dilg et~al.}(1973)\citenamefont{Dilg, Schantl, Vonach,
  and Uhl}}]{DILG1973269}
\bibinfo{author}{\bibfnamefont{W.}~\bibnamefont{Dilg}},
  \bibinfo{author}{\bibfnamefont{W.}~\bibnamefont{Schantl}},
  \bibinfo{author}{\bibfnamefont{H.}~\bibnamefont{Vonach}}, \bibnamefont{and}
  \bibinfo{author}{\bibfnamefont{M.}~\bibnamefont{Uhl}},
  \bibinfo{journal}{{Nucl. Phys. A}} \textbf{\bibinfo{volume}{217}},
  \bibinfo{pages}{269 } (\bibinfo{year}{1973}), ISSN \bibinfo{issn}{0375-9474}.

\bibitem[{\citenamefont{Brink}(1957)}]{BRINK1957215}
\bibinfo{author}{\bibfnamefont{D.~M.} \bibnamefont{Brink}},
  \bibinfo{journal}{{Nucl. Phys.}} \textbf{\bibinfo{volume}{4}},
  \bibinfo{pages}{215 } (\bibinfo{year}{1957}), ISSN \bibinfo{issn}{0029-5582}.

\bibitem[{\citenamefont{Axel}(1962)}]{AxelPhysRev.126.671}
\bibinfo{author}{\bibfnamefont{P.}~\bibnamefont{Axel}},
  \bibinfo{journal}{{Phys. Rev.}} \textbf{\bibinfo{volume}{126}},
  \bibinfo{pages}{671} (\bibinfo{year}{1962}).

\bibitem[{\citenamefont{Ignatyuk et~al.}(1993)\citenamefont{Ignatyuk, Weil,
  Raman, and Kahane}}]{1993Ignatyuk}
\bibinfo{author}{\bibfnamefont{A.~V.} \bibnamefont{Ignatyuk}},
  \bibinfo{author}{\bibfnamefont{J.~L.} \bibnamefont{Weil}},
  \bibinfo{author}{\bibfnamefont{S.}~\bibnamefont{Raman}}, \bibnamefont{and}
  \bibinfo{author}{\bibfnamefont{S.}~\bibnamefont{Kahane}},
  \bibinfo{journal}{{Phys. Rev. C}} \textbf{\bibinfo{volume}{47}},
  \bibinfo{pages}{1504} (\bibinfo{year}{1993}).

\bibitem[{\citenamefont{Capote et~al.}(2009)\citenamefont{Capote, Herman,
  Obložinský, Young, Goriely, Belgya, Ignatyuk, Koning, Hilaire, Plujko
  et~al.}}]{2009Capote}
\bibinfo{author}{\bibfnamefont{R.}~\bibnamefont{Capote}},
  \bibinfo{author}{\bibfnamefont{M.}~\bibnamefont{Herman}},
  \bibinfo{author}{\bibfnamefont{P.}~\bibnamefont{Obložinský}},
  \bibinfo{author}{\bibfnamefont{P.}~\bibnamefont{Young}},
  \bibinfo{author}{\bibfnamefont{S.}~\bibnamefont{Goriely}},
  \bibinfo{author}{\bibfnamefont{T.}~\bibnamefont{Belgya}},
  \bibinfo{author}{\bibfnamefont{A.}~\bibnamefont{Ignatyuk}},
  \bibinfo{author}{\bibfnamefont{A.}~\bibnamefont{Koning}},
  \bibinfo{author}{\bibfnamefont{S.}~\bibnamefont{Hilaire}},
  \bibinfo{author}{\bibfnamefont{V.}~\bibnamefont{Plujko}},
  \bibnamefont{et~al.}, \bibinfo{journal}{{Nucl. Data Sheets}}
  \textbf{\bibinfo{volume}{110}}, \bibinfo{pages}{3107 }
  (\bibinfo{year}{2009}), ISSN \bibinfo{issn}{0090-3752},
  \bibinfo{note}{special Issue on Nuclear Reaction Data}.

\bibitem[{\citenamefont{Goriely et~al.}(2001)\citenamefont{Goriely, Tondeur,
  and Pearson}}]{GORIELY2001311}
\bibinfo{author}{\bibfnamefont{S.}~\bibnamefont{Goriely}},
  \bibinfo{author}{\bibfnamefont{F.}~\bibnamefont{Tondeur}}, \bibnamefont{and}
  \bibinfo{author}{\bibfnamefont{J.~M.} \bibnamefont{Pearson}},
  \bibinfo{journal}{{Atom. Data Nucl. Data}} \textbf{\bibinfo{volume}{77}},
  \bibinfo{pages}{311} (\bibinfo{year}{2001}), ISSN \bibinfo{issn}{0092--640X}.

\bibitem[{\citenamefont{Goriely et~al.}(2008)\citenamefont{Goriely, Hilaire,
  and Koning}}]{Goriely_HilairePhysRevC.78.064307}
\bibinfo{author}{\bibfnamefont{S.}~\bibnamefont{Goriely}},
  \bibinfo{author}{\bibfnamefont{S.}~\bibnamefont{Hilaire}}, \bibnamefont{and}
  \bibinfo{author}{\bibfnamefont{A.~J.} \bibnamefont{Koning}},
  \bibinfo{journal}{{Phys. Rev. C}} \textbf{\bibinfo{volume}{78}},
  \bibinfo{pages}{064307} (\bibinfo{year}{2008}).

\bibitem[{\citenamefont{Goriely}(1998)}]{1998Goriely}
\bibinfo{author}{\bibfnamefont{S.}~\bibnamefont{Goriely}},
  \bibinfo{journal}{{Phys. Lett. B}} \textbf{\bibinfo{volume}{436}},
  \bibinfo{pages}{10 } (\bibinfo{year}{1998}), ISSN \bibinfo{issn}{0370-2693}.

\bibitem[{\citenamefont{Hilaire et~al.}(2012)\citenamefont{Hilaire, Girod,
  Goriely, and Koning}}]{2012Hilaire}
\bibinfo{author}{\bibfnamefont{S.}~\bibnamefont{Hilaire}},
  \bibinfo{author}{\bibfnamefont{M.}~\bibnamefont{Girod}},
  \bibinfo{author}{\bibfnamefont{S.}~\bibnamefont{Goriely}}, \bibnamefont{and}
  \bibinfo{author}{\bibfnamefont{A.~J.} \bibnamefont{Koning}},
  \bibinfo{journal}{{Phys. Rev. C}} \textbf{\bibinfo{volume}{86}},
  \bibinfo{pages}{064317} (\bibinfo{year}{2012}).

\bibitem[{\citenamefont{Arteaga and Ring}(2008)}]{2008Arteaga}
\bibinfo{author}{\bibfnamefont{D.~P.} \bibnamefont{Arteaga}} \bibnamefont{and}
  \bibinfo{author}{\bibfnamefont{P.}~\bibnamefont{Ring}},
  \bibinfo{journal}{Phys. Rev. C} \textbf{\bibinfo{volume}{77}},
  \bibinfo{pages}{034317} (\bibinfo{year}{2008}).

\bibitem[{\citenamefont{Martini et~al.}(2014)\citenamefont{Martini, Hilaire,
  Goriely, Koning, and Péru}}]{2014Martini}
\bibinfo{author}{\bibfnamefont{M.}~\bibnamefont{Martini}},
  \bibinfo{author}{\bibfnamefont{S.}~\bibnamefont{Hilaire}},
  \bibinfo{author}{\bibfnamefont{S.}~\bibnamefont{Goriely}},
  \bibinfo{author}{\bibfnamefont{A.}~\bibnamefont{Koning}}, \bibnamefont{and}
  \bibinfo{author}{\bibfnamefont{S.}~\bibnamefont{Péru}},
  \bibinfo{journal}{Nuclear Data Sheets} \textbf{\bibinfo{volume}{118}},
  \bibinfo{pages}{273 } (\bibinfo{year}{2014}), ISSN \bibinfo{issn}{0090-3752}.

\bibitem[{\citenamefont{Spyrou et~al.}(2007)\citenamefont{Spyrou, Becker,
  Lagoyannis, Harissopulos, and Rolfs}}]{spyrou2007cross}
\bibinfo{author}{\bibfnamefont{A.}~\bibnamefont{Spyrou}},
  \bibinfo{author}{\bibfnamefont{H.-W.} \bibnamefont{Becker}},
  \bibinfo{author}{\bibfnamefont{A.}~\bibnamefont{Lagoyannis}},
  \bibinfo{author}{\bibfnamefont{S.}~\bibnamefont{Harissopulos}},
  \bibnamefont{and} \bibinfo{author}{\bibfnamefont{C.}~\bibnamefont{Rolfs}},
  \bibinfo{journal}{{Phys. Rev. C}} \textbf{\bibinfo{volume}{76}},
  \bibinfo{pages}{015802} (\bibinfo{year}{2007}).

\bibitem[{\citenamefont{Gy\"urky et~al.}(2001)\citenamefont{Gy\"urky, Somorjai,
  F\"ul\"op, Harissopulos, Demetriou, and Rauscher}}]{2001_Gyurky}
\bibinfo{author}{\bibfnamefont{G.}~\bibnamefont{Gy\"urky}},
  \bibinfo{author}{\bibfnamefont{E.}~\bibnamefont{Somorjai}},
  \bibinfo{author}{\bibfnamefont{Z.}~\bibnamefont{F\"ul\"op}},
  \bibinfo{author}{\bibfnamefont{S.}~\bibnamefont{Harissopulos}},
  \bibinfo{author}{\bibfnamefont{P.}~\bibnamefont{Demetriou}},
  \bibnamefont{and} \bibinfo{author}{\bibfnamefont{T.}~\bibnamefont{Rauscher}},
  \bibinfo{journal}{Phys. Rev. C} \textbf{\bibinfo{volume}{64}},
  \bibinfo{pages}{065803} (\bibinfo{year}{2001}).

\end{thebibliography}

\end{document}